\documentclass[aps,prx,reprint,groupedaddress,a4paper]{revtex4-1}

\usepackage{mathtools,bm,amsmath,amssymb}
\usepackage{letltxmacro}
\LetLtxMacro{\originaleqref}{\eqref}
\renewcommand{\eqref}{Eq.~\originaleqref}
\newcommand{\eq}{\begin{equation}}
\newcommand{\qe}{\end{equation}}
\newcommand{\dd}[2]{\frac{d#1}{d#2}}
\newcommand{\pdd}[2]{\frac{\partial#1}{\partial#2}}
\newcommand{\non}{\nonumber}
\usepackage{adjustbox}
\DeclareMathOperator{\csch}{csch}

\begin{document}

\title{Pattern selection in reaction diffusion systems}

\author{Srikanth Subramanian}

\author{Se\'an M. Murray}
\email{sean.murray@synmikro.mpi-marburg.mpg.de}
\affiliation{Max Planck Institute for Terrestrial Microbiology, Marburg, Germany}

\begin{abstract}
Turing's theory of pattern formation has been used to describe the formation of self-organised periodic patterns in many biological, chemical and physical systems. However, the use of such models is hindered by our inability to predict, in general, which pattern is obtained from a given set of model parameters. While much is known near the onset of the spatial instability, the mechanisms underlying pattern selection and dynamics away from onset are much less understood. Here, we provide new physical insight into the dynamics of these systems. We find that peaks in a Turing pattern behave as point sinks, the dynamics of which are determined by the diffusive fluxes into them.  As a result, peaks move towards a periodic steady-state configuration that minimizes the mass of the diffusive species. We also show that the preferred number of peaks at the final steady-state is such that this mass is minimised. Our work presents mass minimization as a simple, generalisable, physical principle for understanding pattern formation in reaction diffusion systems far from onset.
\end{abstract}

\maketitle

\section{Introduction}
Pattern formation occurs in a huge variety of natural and living systems\label{intro} \cite{Ball1999}, from chemical reactions \cite{Kapral1995,Kuramoto2003} to living cells \cite{Kondo2010,Koch1994,Halatek2018} to environmental patterns \cite{Meron2019}. In systems described by reaction-diffusion (RD) equations, the formation of spatially periodic patterns can be explained by the Turing instability, in which patterns emerge due to the presence of two or more interacting components that diffuse (or are transported) at different rates \cite{Turing1952, Murray2008, Cross2009, Maini2012}. The resulting patterns are multi-stable in that several different stable patterns can be obtained from the same set of parameters, albeit, for incompletely understood reasons, with different frequencies \cite{Murray2017}.

Sufficient conditions for pattern formation can be determined in the so-called Turing or linear regime, in which a spatially uniform stable steady state becomes linearly unstable to spatial perturbations in the presence of diffusion \cite{Murray2008}. 
Consider the following one-dimensional system
\begin{subequations}
\begin{align}
\partial_t u &= D_u \partial^2_x u +f(u,v)\\
\partial_t v &= D_v \partial^2_x v +g(u,v)~.
\end{align}
\end{subequations}
The evolution of any small perturbation from a spatially uniform steady state is given by its decomposition into its Fourier modes $e^{\sigma_k t} cos(kx)$, where $Re(\sigma_k)$ is the growth rate. Then, the uniform steady state is laterally unstable if any wave number $k$ has a positive growth rate $Re(\sigma_k)>0$ (see Fig. 1A and Supporting Information). For a finite domain $[0,L]$ and reflexive boundary conditions, the wave number $k$ is discrete with $k=\frac{n\pi}{L}$ for integer $n$. The unstable modes grow exponentially in time until the non-linear terms can no longer be neglected. These terms saturate the exponential growth and select different spatial states.
At the onset of the instability when a single mode $n_c$ is unstable, the naive expectation is that the growth saturates without substantially changing the spatial structure so that the final sate is qualitatively similar to mode $n_c$. This saturation is described by the corresponding amplitude equations \cite{Cross1993,Cross2009,DeWit1999}.
These equations and, more generally, the weakly non-linear approach on which they are based, have been extremely useful in understanding the selection and stability of different fundamental modes in a variety of systems \cite{DeWit1999,Cross1993,Cross2009,Pena2000,Pena2001,Cates2010}, as well as the effect of external constraints such as fixed boundary conditions, parameter ramps, external forcing, template patterns, system geometry and deformable or moving boundaries \cite{Arcuri1986,Bergmann2018,Rapp2016,Ruppert2020, Bergmann2019,Wigbers2020, Vanag2003, Dolnik2001,Pena2003,Miguez2004,Yang2006}.

However the approach is only valid close to onset (in the vicinity of the bifurcation). Away from onset, where many modes are linearly unstable (Fig 1A), pattern forming RD systems typically still produce patterns with a well defined periodicity, corresponding to (in 1D) a particular mode number $n$ and its harmonics $2n, 3n, \ldots$ (Fig. 1B). This is despite neighbouring modes $n\pm1$ generally having similar growth rates, which would be expected to lead to aperiodic patterns. The physics underlying this `exclusion principle' \cite{Arcuri1986} are in general not known\footnote{See section 9.1.4 of \cite{Cross2009} for a more detailed discussion.}. This is very relevant as non-equilibrium systems in nature cannot be expected to be close to onset.

In the following, we propose a simple physical principle to explain the dynamics, positioning and number of peaks in a Turing pattern far from onset. Inspired by our previous work \cite{Murray2017}, we begin by reviewing a number of observations about patterns on a 1D domain. As already mentioned final patterns away from onset still have a well-defined wavelength. With reflexive boundary conditions, the peaks are also regularly positioned i.e. the peaks (or valleys) of the pattern are found at the same locations as those of some fundamental mode $n$. While this regular positioning is consistent with the selection of a particular mode, it appears, for the reasons given above, that this is a non-linear effect.

We will also see below that if a system is initialised with a mis-positioned pattern (and therefore far outside of the linear regime), for example a single mis-positioned peak, then the peak subsequently moves towards mid-domain without substantially changing its shape. This is also evident in models that exhibit coarsening \cite{Murray2017, Brauns2020,Tzou2013,Kolokolnikov2005,Kolokolnikov2006}, which we define here as the preference for a steady state pattern dominated by a mode lower than that predicted by linear stability i.e. the mode with greatest linear growth rate $\sigma$. In Fig 1C, we show the evolution of a pattern starting from a small perturbation of the uniform state. The pattern initially resembles mode $n=7$ (three and a half peaks) consistent with the linear prediction (Fig. 1A) but it subsequently coarsens, first to three peaks and then to two. After each coarsening event, the peaks move towards their regularly positioned configuration so that the final steady-state pattern consists of a peak at each quarter position (mode 4). In two-variable mass-conserved systems, coarsening is complete in that, irrespective of how many peaks there are initially, the pattern eventually coarsens down to a single peak or half-peak (monotonic) solution \cite{Otsuji2007,Ishihara2007,Morita2010, Morita2012, Jimbo2013}. On a periodic domain, absolute positioning is no longer meaningful but peaks still re-position to maintain a constant wavelength. Finally, regular positioning is maintained even during domain growth in which new peaks are created by insertion or splitting \cite{Crampin1999,Maini2012, Murray2017}. Overall, these observations indicate that the periodic positioning of peaks is an inherently non-linear effect and not a remnant or direct consequence of the dominating linear mode of the base state perturbation. Thus, while a decomposition into fundamental modes is critical to understanding the initial formation of the pattern (starting from the homogeneous state), once peaks have formed, a different description is required.

Note also that the two phases of a pattern that exist when imposing reflexive boundary conditions are not necessarily equally preferred. We have previously studied a model in which the pattern consisting of a single peak at mid-domain is preferred over a half-peak at each boundary and similarly for higher modes \cite{Murray2017}. Thus, not only is the mode of the pattern selected, the phase is too. However, peaks on the boundary display different dynamics: unlike interior peaks, they do not move but only appear or disappear. Here, we will restrict ourselves to the study of interior peaks only as they are amenable to comparison with point sinks and hence we will not address the issue of phase selection. We make this explicit in the last section by using periodic boundary conditions. In the interim, we will use reflexive boundary conditions in order to more easily study peak movement.

In the following, we show that the peaks of a Turing pattern behave as point sinks that move with a velocity proportional to the diffusive flux across them. This is a consequence of the flow of mass through the system is responsible for the regular positioning of peaks. By flow, we mean something more than simply the flux through the system. In a diffusive non-mass-conserving system, `molecules' enter the system, diffuse and either leave the system or are converted to another species. This combination of diffusion and turnover results, as we shall see, in the regular positioning of peaks due to the concept of flux-balance \cite{Sugawara2011, Ietswaart2014}. This result also explains why the peaks in mass-conserving two-variable reaction-diffusion systems do not move: there is no flow to drive the movement.

We also find that the regularly positioned configuration minimises the total mass of the rapidly diffusing species, the substrate of the nonlinear reaction. We then find empirically that this `minimisation principle' can be extended to predict not only the final positions of the peaks but also the final {\it number} of peaks, even in the presence of coarsening. This is significant as the amplitude equation approach for determining the dominant mode is not applicable far from onset. The principle of mass minimisation is therefore an incredibly simple yet powerful concept for understanding the behaviour of pattern-forming systems.

\begin{figure}[htb]
 \centering
   \includegraphics[width=\linewidth]{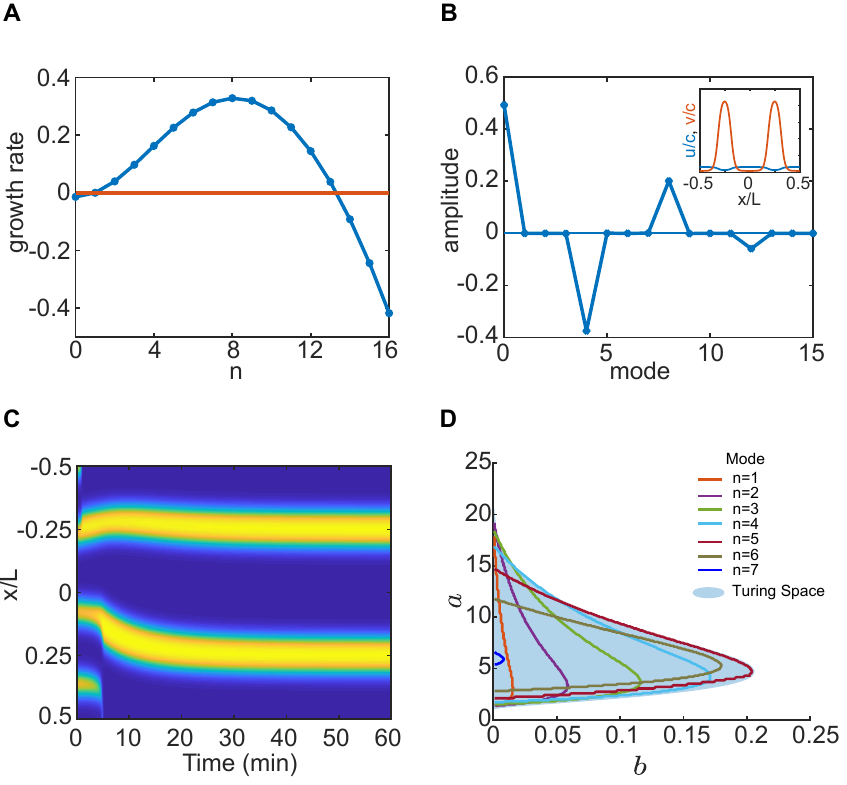}
 \caption{The Turing instability of reaction-diffusion systems.
 (\textit{A}) The growth rate of different modes for the model in \eqref{eq:model_equations} with default parameters and $L=4$. Note that the growth rate at $n=0$ is negative -- the system is not generically mass conserving. (\textit{B}) The Fourier decomposition of the obtained two-peak pattern (inset). (\textit{C}) An example kymograph showing pattern development starting from a random perturbation of the uniform state. While mode $n=7$ dominates initially, the pattern coarsens down to two peaks, dominated by mode $n=4$. See also Fig. S1. (\textit{D}) The regions of instability of each mode on a domain of length $L=2$ (the region bounded by the y-axis and the corresponding coloured curve). The blue shaded region shows the Turing space for an infinite domain (see also Fig. S1 and the Supporting Information for further details).}
 \label{fig1}
\end{figure}

\section*{The model}
We introduce the following exploratory one-dimensional system, inspired by our recent model of bacterial condensin \cite{Murray2017,Hofmann2019}, written in terms of the variables $u=u(x,t)$ and $v=v(x,t)$,
\begin{subequations}
\begin{align}
\partial_t u &= D_u \partial^2_x u -\beta u(u+v)^2 + \gamma v + c\delta -\delta u\\
\partial_t v &= D_v \partial^2_x v +\beta u(u+v)^2 - \gamma v -\delta v~,\label{eq:model_equation_v}
\end{align}
\label{eq:model_equations}
\end{subequations}
defined over the spatial domain $[-L/2,L/2]$, with reflexive boundary conditions, all parameters non-negative and $D_v<D_u$. While superficially similar to the some of the classic Turing models such as the Brusselator \cite{Prigogine1968} and Schnakenberg \cite{Schnakenberg1979} models, this model has some notable properties that make some analyses easier. In the absence of diffusion, it has a single fixed point that is stable for all parameter values. This means that the stability diagram of the system is particularly simple. There are only two regions, specified by a single inequality: one in which the spatially uniform solution is stable and another in which it is Turing unstable (Fig. 1D). There are no oscillatory instabilities. 
Like the Brusselator, the model has the form of a mass-conserving Turing system with additional terms: a global source term, $c\delta$, and two depletion terms, $\delta u$ and $\delta v$. By writing the source term as $c\delta$, we can change $\delta$, the turnover rate, while leaving the total steady state concentration $c$ fixed. We obtain a mass-conserved Turing model when $\delta=0$ and the limit $\delta\rightarrow 0$ is well defined as long as we constrain the total initial mass to be the same as the steady-state mass, i.e. $C(0)=c$. 

The condition for a Turing instability is most easily stated by non-dimensionalising the system and introducing the dimensionless parameters  $a=\frac{\beta c^2}{\gamma},~b=\frac{\delta}{\gamma},~\Gamma=\frac{\gamma L^2}{D_v},~d=\frac{D_u}{D_v}$ (see Supporting Information for details). As can be seen in Figure 1D for typically choices of the diffusivity ratio $d$, we require $b\ll 1$ for patterning, i.e. the timescale of mass flow (turnover) through system, $1/\delta$, must be much longer than the timescale underlying the Turing instability $1/\gamma$.

Numerically solving the system, we found that it indeed produces regularly positioned peaks. We also observed that, like the model it is based on \cite{Murray2017}, it exhibits a competition instability \cite{Tzou2013,Kolokolnikov2005, Kolokolnikov2006} (also known as interrupted coarsening \cite{Brauns2020}) in that the final dominant mode has a shorter wavelength than predicted by linear stability analysis. For our default parameter set with $L=4$ ($\Gamma=4800$), linear stability predicts (Fig. 1A) that the pattern consists of four peaks (or valleys) (mode $n=8$) whereas the obtained steady-state pattern most frequently consists of two peaks (mode $n=4$) (Fig. 1B, Fig. S1D). While multiple peaks often form initially, consistent with the linear prediction, coarsening rapidly occurs, leaving mis-positioned peaks that then move slowly towards opposite quarter positions, while maintaining their shape (Fig. 1C). Note that this movement is only observed because of the competition instability. It is not evident in models/parameters sets for which the linear prediction holds as in that case, the peaks are created at their steady-state positions. We will return to this incomplete coarsening later.

To examine the movement of peaks in more detail, we focused on the case of a single peak ($n=2$), typically obtained for $L=2$ ($\Gamma=1200$). Examining the movement of the peak (Fig. 2A), we found that it moves to mid-domain exponentially in time (Fig. 2B), indicating the peak velocity is linearly proportional to its displacement from mid-domain (Fig. 2B, bottom inset). This was the case whether the system was initialised with a random perturbation of the uniform state or with a peak preformed somewhere on the domain. That peaks might move to respect the symmetry of the system is perhaps, while underappreciated, not surprising though it is relevant for understanding the periodicity and positioning of peaks. Indeed, the rate of movement was found to be directly proportional to the turnover rate $\delta$ (Fig. 2B, top inset) (or equivalently $c\delta$ the flux through the system per unit length) so that in the mass-conserved limit, $\delta\rightarrow0$, peaks do not move (Fig. 2B,C). This is consistent with our previous results \cite{Murray2017}.

\begin{figure}[tb]
 \centering
   \includegraphics[width=\linewidth]{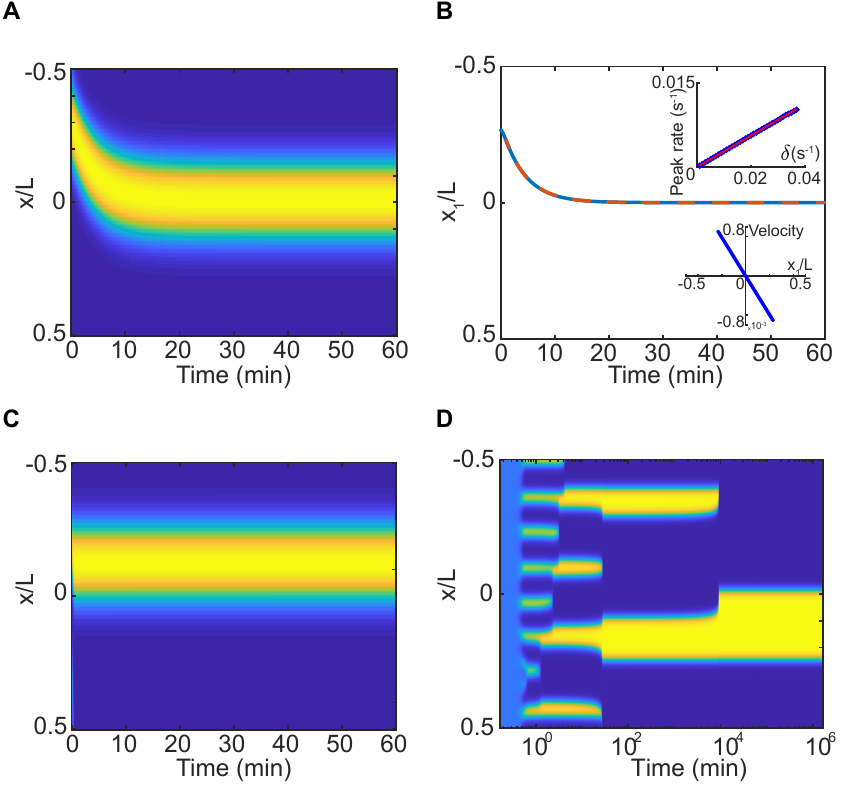}
 \caption{Peak movement and regular positioning depend on flux through the system. (\textit{A}) The system is initialised with a peak away from mid-domain. The peak subsequently moves to mid-domain. (\textit{B}) The centroid of the peak (blue line) plotted as function of simulation time. The orange dashed line is an exponential fit. Inset: (Top right) The rate of movement obtained from fitting the centroid to an exponential as in (\textit{B}) shows a linear dependence on the turnover rate $\delta$. (Bottom right) Peak velocity is linear in peak position. (\textit{C}) A single peak in the mass conserved limit $\delta=0$ can be positioned anywhere on the domain. No peak movement is observed (\textit{D}) The mass conserved system exhibits complete coarsening. Irrespective of how many peaks there are initially, the pattern eventually coarsens to a single peak, the position of which depends on which peak of the initial pattern has not coarsened. In (\textit{D}) $\Gamma=19200$ ($L=10)$.
 }
 \label{fig2}
\end{figure}
Mass-conserving RD models exhibit a complete coarsening process in that the final steady-state pattern is either mono-modal (periodic or reflexive) or monotonic (reflexive only) depending on the boundary conditions, as has been proved explicitly for several models \cite{Otsuji2007,Ishihara2007,Morita2010,Morita2012,Jimbo2013}. We find the same coarsening behaviour here (Fig. 2D). We only obtain the half-peak solution for very short domains i.e. when the width of interface is comparable to the domain length. If the domain length or other parameters are chosen such that there is initially more than one peak then the coarsening process results (eventually) in a single interior peak (Fig. 2D). Importantly, since peaks do not move, the position of this final peak is determined by whichever peak of the transient state remains after coarsening i.e. the steady-state solutions are in general not symmetric as might naively be expected by the boundary conditions. We tested these conclusions by initialising the system with a single preformed peak (constructed as a translation of the non-mass conserved steady-sate solution). We found that preformed peaks do not move and constitute a stable solution (Fig. 2C). Thus, the mass-conserved case $b=0$ with reflexive boundary conditions has a continuum of single-peak stable states, whereas there is at most one unique single-peak solution for $b>0$. This implies that regular positioning is not an intrinsic property of the system but rather depends on $b$. These results are based on simulations that were run for very long times with very low error tolerances and are in agreement with our previous results \cite{Murray2017}. We will also see the same behaviour when we consider point sinks in the next section. Overall, these results demonstrate a connection between peak movement towards the regular positioned configuration and the flow (turnover) of mass through the system.

\section*{Point sinks}
To explore this connection in more detail, we turn to a toy model involving diffusion and point sinks. We consider the steady-state diffusion equation for a variable $A=A(x)$ over a one-dimensional domain of length $L$ in the presence of global source and decay terms as well as $n$ localised point sinks at positions $\bm{x}=(x_1,\cdots,x_n)$ (each with rate $\mu$):
\eq  D\frac{d^2A}{dx^2} + c\delta - \delta A - \sum_{i=1}^n \mu L \delta (x-x_i)A =0 \label{eq: plasmid_eq} ~.\qe
We take the domain to be $[-L/2,L/2]$ and impose zero-flux boundary conditions. As before, we write the global source term in terms of the decay rate $\delta$ and a concentration $c$, which is the steady-state concentration in the absence of the point sinks.
A simpler system without the decay term and with perfect points sinks (i.e. $\mu\rightarrow \infty$) was used by Ietswaart et al. to model the positioning of plasmids within rod-shaped bacterial cells \cite{Ietswaart2014}. They found that the gradient differential across each sink vanishes if and only if the sinks are regularly positioned and, therefore, if sinks were to move up the concentration gradient, they would be regularly positioned. We will extend this result to the more complicated case of equation \eqref{eq: plasmid_eq}. Note that the presence of the decay term introduces an additional length scale $\sqrt{D/\delta}$ into the system, namely the distance that a molecule of $A$ would diffuse (in the absence of any point sinks) before it decays. We refer to this as the length-scale of diffusion. It is small when either diffusion is slow or the decay rate (turnover) $\delta$ is fast.

 We can write the solution to \eqref{eq: plasmid_eq} as
 \eq A(x) = c-\sum_i \mu'_i G(x;x_i)~, \label{eq: plasmid_sol}\qe
 where $G(x;x_i)$ is the modified Green's function defined by
\begin{align}
\begin{split}
-\frac{L^2}{\kappa^2}G_{xx} (x;x_i) + G(x;x_i) = L\delta (x-x_i) \\
G_x(\pm \frac{L}{2}; x_i) = 0,\quad \frac{1}{L}\int_{-\frac{L}{2}}^{\frac{L}{2}} G(x;x_i) dx = 1~,
\end{split}
\label{eq:greens_function_equation}
\end{align}
where the dimensionless parameter $\kappa=L\sqrt{\frac{\delta}{D}}$ is the ratio of the length of domain to the length-scale of diffusion.
The coefficients $\mu'_i=\mu'_i(\bm{x})$ are determined by the linear algebraic conditions
\eq \mu'_i=\lambda A(x_i)\quad i=1,\ldots,n~,\label{eq:linear_system}\qe
where we have defined a second dimensionless parameter $\lambda=\frac{\mu}{\delta}$, the ratio of the sink and background decay rates.
The quantities $\mu'_i$ have a simple interpretation.
They are directly related to $J_i$, the flux leaving the system through each sink
\begin{align}
 J_i &=J_{i+}+J_{i-}\non\\
     &= -D\sum_j \mu'_j \left[ G_x(x_i^+;x_j) - G_x(x_i^-;x_j)\right]=L\delta\mu'_i~,\non
\end{align}
where $J_i=|D\dd{A}{x}|$ and the $-$ and $+$ subscripts refer to the diffusive flux from the left and right respectively. 
We also define the flux differential across each sink as
\begin{align}
\Delta J_i &= \frac{1}{2}(J_{i+}-J_{i-}) \non\\
    &= -\frac{D}{2}\sum_j \mu'_j \left[ G_x(x_i^+;x_j) + G_x(x_i^-;x_j) \right]~.
    \label{eq:flux_differential_plasmid}
\end{align}
Note the total mass (concentration) of $A$ in the system is readily given by
\eq M\coloneqq\frac{1}{L}\int_{-\frac{L}{2}}^{\frac{L}{2}} A(x) dx = c - \sum_i \mu'_i 
\qe
where the second term solely describes the effect of the point sinks.

We can now investigate what would in happen in this system if sinks were to move up the gradient of $A$. As in Ietwaart et al., we can determine the configurations for which the flux differentials are all zero. In Appendix A, we prove that this occurs uniquely for regularly positioned sinks, $x_i=\bar{x}_i :=(i-\frac{1}{2})\frac{L}{n}-\frac{L}{2}$, i.e.
\eq \Delta J_i(\bm{\bar{x}})=0 \quad \textrm{for all }i~.\non\qe
Interestingly, we also show that the regular positioned configuration $\bar{x}_i$ is the unique stationary point of the mass $M$, i.e.
\eq \left. \pdd{}{x_i} M(\bm{x})\right|_{\bm{x}=\bm{\bar{x}}}=0 \quad \textrm{for all }i.\non\qe

Thus if sinks move up the concentration gradient (in the direction of greatest flux), they will be regularly positioned as this is the configuration for which the fluxes into each sink from either side balance. Furthermore this configuration minimizes the total mass of the system. In other words the sinks are positioned so as to `consume' mass at the greatest rate. This connection between regular positioning and mass minimisation appears to be generalisable and we have observed numerically that it holds for spatial sinks i.e. if the delta function in equation \eqref{eq: plasmid_eq} is replaced by a peak-shaped spatial function such as a Gaussian function or  $\mathrm{sech}^2(x)$, then the total mass is minimised when the sink is centred at mid-domain.

\begin{figure}[tb]
\centering
 \includegraphics[width=\linewidth]{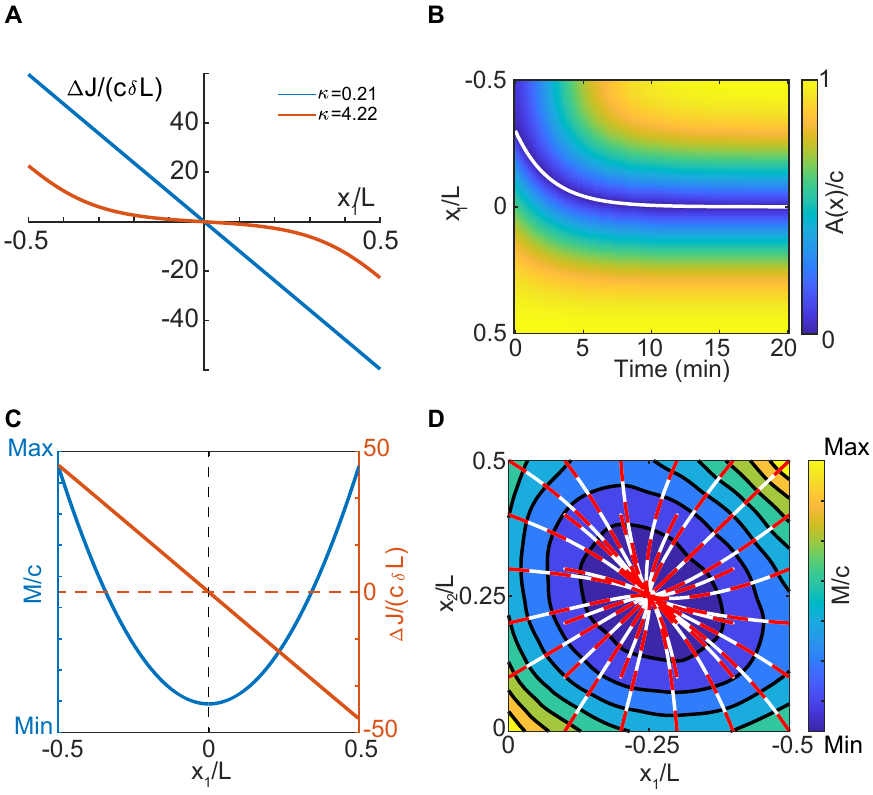}
 \caption{Moving point sink are regularly positioned and their movement depends on the diffusive length-scale. (\textit{A}) Flux differential across a point sink calculated analytically as a function of sink position $x_1$ for two values of $\kappa$. (\textit{B}) When $\Delta J$ is linear in $x_1$ the sink moves exponentially to mid-domain. The flux differential across a point sink is linear in sink position for $\kappa \ll 1$. It vanishes at the middle of the domain. (\textit{C}) Mass $M$ in system as a function of sink position for a single point sink is plotted in blue. The Mass is minimal as the sink approaches the middle of the domain. (\textit{D}) Sample trajectories of the two-sink system. White lines are sample trajectories obtained using \eqref{eq:sink_velocity}, while the overlaid red dashed lines are trajectories obtained using \eqref{eq:sink_velocity_mass}. The coloured contour shows the total mass $M$ as a shown as a function of the sink positions. The minimum occurs at the steady-state configuration (sinks at opposite quarter positions). Parameters: $D = 0.3, \lambda = 166.1, c=1, L=1, \nu = 1$. $\kappa=0.21$ in (B) and $4.22$ in (\textit{C}). In (\textit{D}) $L=2$. }
\label{fig3}
\end{figure}

Let us consider the case of a single sink, $n=1$, in more detail. We focus on the regime $\kappa\ll 1$ in which the diffusive length-scale is much longer than the domain size. We expand in $\kappa$ to find first
\eq \frac{\mu'_1}{c}\approx\frac{\lambda}{\lambda+1}-\frac{\lambda^2}{\lambda+1}\left(\frac{x_1^2}{L^2}+\frac{1}{12}\right)\kappa^2+O(\kappa^4)\non\qe
and then
\begin{align} \frac{\Delta J_1}{c\delta L}&=-\frac{1}{2}\frac{\mu'_1}{c}\frac{\sinh(2\kappa\frac{x_1}{L})}{\sinh(\kappa)}\non\\
&\approx-\frac{\lambda}{\lambda+1}\frac{x_1}{L}+O(\kappa^2)~.\label{eq:flux_one_sink}
\end{align}
Hence, if $\kappa\ll 1$, then the flux differential across the sink depends linearly on its relative displacement from mid-domain. For strong sinks ($\lambda \gg 1$), the proportionality factor is linear in $\delta$, just as we observed for the Turing system (Fig. 2B). As $\kappa$ increases, the flux-differential becomes inflected about $x_1=0$ (Fig. 3A). We can think of this heuristically as follows. If the diffusive length-scale is much shorter than the domain size ($\kappa \gg 1$), then only particles initially created near the sink will fall into it. Therefore the flux-differential is only significantly non-zero close to the boundaries (or another sink). In essence, the geometry sensing of the system breaks down. On the other hand, when the diffusive length-scale is much longer than the domain size ($\kappa \ll 1$), particles can explore the entire domain before decaying and so the flux-differential across the sink reflects its position on the domain, with the fluxes into the sink from either side balancing at mid-domain. The relevance of this dependence on $\kappa$ to Turing systems will be made clear later.
 
We can make sink movement explicit by specifying the sink velocities. Given our results above, two natural choices are to take the sink velocities as either directly proportional to the flux-differentials $\Delta J_i$,
\begin{align}
\dd{x_i}{t}&=\nu \Delta J_i(\bm{x}) ~\label{eq:sink_velocity}
\end{align}
or to the derivative of the mass $M(\bm{x})$ with respect to the sink position
\begin{align}
\dd{x_i}{t}=-\frac{n}{2}\nu D\pdd{}{x_i}M(\bm{x}) ~,\label{eq:sink_velocity_mass}
\end{align}
where $\nu$ is some parameter. For the latter choice, the system is analogous to that of $n$ over-damped particles moving in a potential $U(\bm{x})/k_B T=\frac{n\nu}{2} M(\bm{x})$. Note also that while the velocities in equation \eqref{eq:sink_velocity} are specified in terms of local quantities, in equation \eqref{eq:sink_velocity_mass}, they are specified in terms of the global quantity, $M(\bm{x})$.

In either case, the steady-state solution consists of regularly positioned sinks as this is the configuration for which the fluxes balance and for which the mass is at its unique minimum. This holds as long as $\kappa>0$, or equivalently $\delta>0$. For $\delta\rightarrow0$, all the velocities vanish identically. Hence all sink positions are stable in that limit. This is the same singular behaviour that we found in the previous section for the mass-conserved Turing system.

If we assume that sinks move on a much slower timescale than that of diffusion, we can use the steady state solution for $A(x)$ given in equation \eqref{eq: plasmid_sol} to solve the dynamic system \eqref{eq:sink_velocity}. We find, as expected, that a single sink moves exponentially to mid-domain (Fig. 3B). We also find that increasing $\delta$, which shortens the diffusive length-scale while also increasing the flux through the system, leads to faster sink movement (Fig. S2) reminiscent of the Turing system (Fig. 2B). On the other hand, if we decrease $D$, which decreases the diffusive length-scale without affecting the flux through the system, the sink moves more slowly towards mid-domain (Fig. S2). We also considered the system with two sinks and confirmed that the steady state solution consists of quarter-positioned sinks, the configuration that the minimizes the total mass of $A$ (Fig. 3D).

We can also use the steady state solution of $A(x)$ to make explicit a correspondence between the two choices for the sink velocities. While the steady-states of the two systems are identical, their dynamics are not in general the same. However, we found that (Appendix B), in the regime of a long diffusive length scale ($\kappa \ll 1$) and strong sinks ($\lambda \gg 1$), the two expressions become equivalent. This equivalence was apparent even for our default parameter set - the sink trajectories arising for either choice were almost identical (Fig. 3D).

\section*{Comparison with the Turing system}

The similarity between moving point-sinks (Fig. 3) and the movement of peaks in a Turing pattern (Figs. 1 and 2) is striking. It suggests that the movement and steady-state positions of peaks in a Turing pattern may be due to a dependence of the peak velocity on the flux-differential (of the fast species across a peak of the slow species) or due to the total mass of the fast species acting as a potential energy surface. Note that in the following we restrict ourselves to Turing patterns consisting only of interior peaks, as boundary peaks are not amenable to a point sink approximation (see below). First, we introduce the following definition of the flux-differential into the peak of a single-peak Turing pattern:
\eq \Delta J_s(t)=D_u\frac{\int_{-L/2}^{L/2}\pdd{u(x,t)}{x}v(x,t) dx}{\int_{-L/2}^{L/2}v(x,t) dx} ~.
\label{eq: flux_differential_peak}
\qe
This is similar to the definition used in a recent spatially-extended model of plasmid positioning \cite{Walter2017}. We initialised the system with a single peak and monitored $\Delta J_s$ as a function of the peak position and velocity. We found that, like for point sinks (Fig. 3C), the flux-differential is, away from the domain boundaries, directly proportional to the displacement from mid-domain (Fig. 4A). Thus, peaks do indeed move with a velocity proportional to the flux-differential. 

However, it is not clear how to extend the definition of the flux-differential to patterns with multiple peaks as well as to higher dimensions in which Turing patterns can consist of complex structures such as stripes, spirals and hexagons. The concept of mass minimisation on the other hand is easy to generalise. When we examined the total mass (concentration) of $u$ (the fast species) in the system, $M=\frac{1}{L}\int_{-\frac{L}{2}}^{\frac{L}{2}} u(x,t)~dx$, we found that it decreases monotonically as the peak moves to mid-domain, modulo boundary effects (Fig. 4A). Further, when we initialised the system with two peaks positioned at various locations, we found similar behaviour (Fig. 4C) suggesting that, for a given number of peaks, the regularly positioned configuration minimises the total mass of $u$, just as we have proven for point sinks in the previous section. Indeed, the trajectories show a remarkable similarity to those of moving point sinks (Fig. 3D).

\begin{figure}[tb]
\centering
\includegraphics[width=\linewidth]{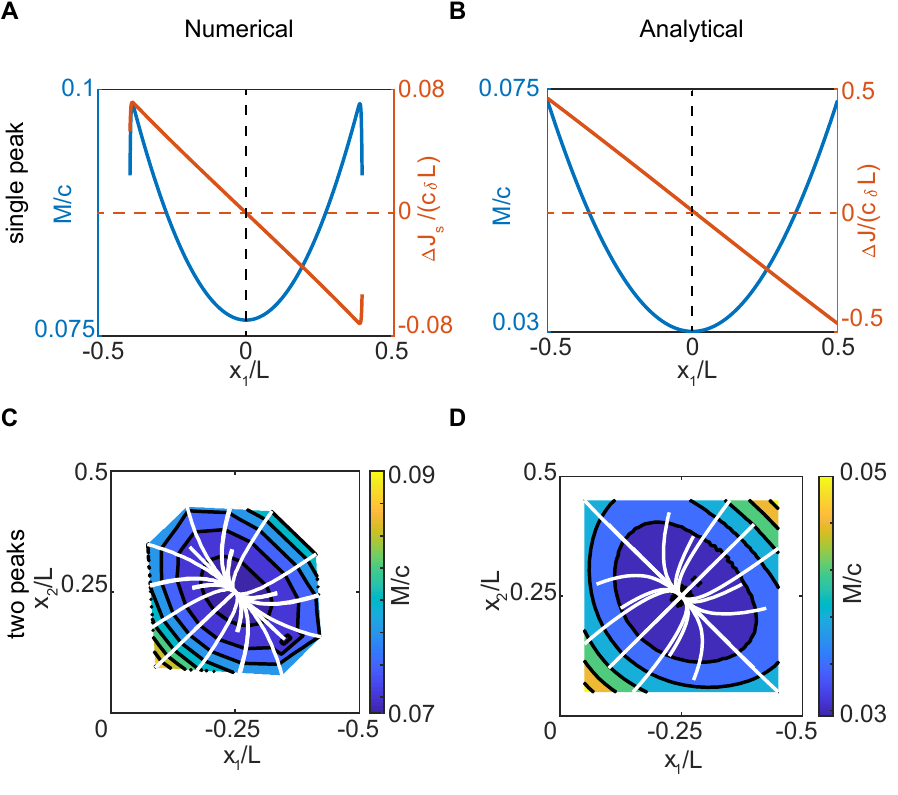}
 \caption{The mass of $u$ is minimised at regular positions. 
(\textit{A}) Flux differential measured numerically using equation \eqref{eq: flux_differential_peak} for a single spike (orange) is a linear function of the peak position. The mass of the fast species $M$ (blue) is minimised at mid-domain. See also Fig. S3 (\textit{B}) The same quantities as in (\textit{A}) but for the analytical expressions from the spike approximation (eq. \eqref{eq: flux_differential_spike} and $M=c-\rho'_{1+}$). (\textit{D}) Mass minimisation for two peaks. Trajectories of two peaks as they move towards opposite quarter positions (while lines). The contours and colour bar represent the mass $M$ interpolated from trajectories. The mass is minimised for regular positioning. (\textit{D}) Same as (\textit{C}) but trajectories obtained from the approximation of peaks as spikes using equation \eqref{eq:velocity_spikes}. Parameters: $L=2$ in (\textit{A}) and (\textit{B}), $L=4$ in (\textit{C}) and (\textit{D}); $D_v=0.0012$, otherwise default. This gives $\sigma=0.0146$ in (\textit{B}) and $\sigma=0.0073$ in (\textit{D})}
 \label{fig4}
\end{figure}

To explore this analytically, we considered the singular limit $D_v \ll D_u$ in which the peaks in $v$ take the form of narrow spikes or pulses of width $\epsilon =\mathcal{O}(\sqrt{D_v/\gamma})$ (Fig. S3). Away from the spike $v$ is approximately constant with a value $v_{out}$ that is much smaller than $u$. This limit allows the use of non-linear analysis methods to study the existence, stability and dynamics of Turing patterns \cite{Wei2014}. Here, our goal is simply to derive an approximation for $u$ in this limit by treating the spikes of $v$ as Dirac delta functions as described below.

We look for steady-state solutions consisting of $n$ spikes at positions $x_1,\cdots,x_n$. 
We assume that $u$ changes slowly within each spike and so can be approximated by a constant $u_i$ and within each spike $u_i \ll v$. 
First we introduce the inner coordinate, $y_i=(x-x_i)/\epsilon$, within each spike. 
We then have the following system for the inner variable $v_i(y)$
\begin{align}
\frac{D_v}{\epsilon^2}\dd{{}^2 v_i}{y_i^2}+\beta u_i v_i^2-(\gamma+\delta)v_i=0~\non\\
v_i\rightarrow 0 \quad \mathrm{as}\quad y_i\rightarrow \pm \infty~,\non
\end{align}
which gives
\begin{align}
v_i=\frac{3}{2}\frac{\gamma+\delta}{\beta u_i}\mathrm{sech}^2(\sqrt{\frac{\gamma+\delta}{D_v}}\frac{\epsilon y_i}{2})~.\non
\end{align}
In the outer region, each spike is approximated by a weighted Dirac delta function and we therefore replace the $v$ and $uv^2$ terms by Dirac delta functions with weights $w_1$ and $w_2$ given by
\begin{align}
w_1 =\epsilon\int_{-\infty}^\infty v_i(y_i) dy_i=6\frac{\sqrt{D_v(\gamma+\delta)}}{\beta u_i}~
\non\\
w_2= \epsilon u_i\int_{-\infty}^\infty v_i^2(y_i) dy_i=6\frac{\sqrt{D_v}(\gamma+\delta)^{3/2}}{\beta^2 u_i}~\non
\end{align}
respectively. 
The equation for $u$ in the outer region then becomes
\begin{align}
D_u\dd{{}^2 u}{x^2} +c\delta -\delta u -\sum_{i=1}^n \frac{\rho}{u}  L\delta(x-x_i)=0\label{eq:outer_equation}
\end{align}
with $\rho=6\frac{\sqrt{D_v}}{L}\frac{\delta\sqrt{\gamma+\delta}}{\beta}$ and where we have used that, according to the spike approximation, $\beta u(u+v_{out})^2-(\gamma+\delta) v_{out} \approx 0$ (from equation \eqref{eq:model_equation_v}) to simplify the contribution away from the spike. Note the inverse dependence on $u$ in the point sink term (which we call an inverted sink). This form is also obtained for other Turing systems with a $uv^2$ non-linearity, such as the Schnakenberg and Brusselator models \cite{Iron2004,Tzou2013}. 

Following the approach of the previous section, the solution to equation \eqref{eq:outer_equation} is given by
\eq u(x) = c- \sum_i \rho'_i G(x;x_i)~, \label{eq:inverse_delta}\qe
where the Green's function is defined as for point sinks but in terms of the corresponding dimensionless parameter $\kappa=L\sqrt{\frac{\delta}{D_u}}$, the ratio of the length of the domain to the diffusive length scale of $u$ (henceforth $\kappa$ replaces $b$ in the set of dimensionless parameters of the system).
The coefficients $\rho'_i=\rho'_i(\bm{x})$ are now determined by the non-linear algebraic system
\eq \rho'_i=\sigma\frac{c^2}{u(x_i)}\quad i=1,\ldots,n~.\label{eq:inverted_algeraic_system}\qe
where $\sigma = \frac{\rho}{c^2 \delta}=6\frac{\sqrt{b+1}}{a\sqrt{\Gamma}}$ is the second dimensionless parameter of \eqref{eq:outer_equation}.
The inverse dependence on $u(x_i)$ makes solving this algebraic system challenging. For a general choice of sink positions $x_i$, there are $n$ coupled quadratic equations in $\rho'_i$, and therefore up to $2^n$ real solutions. However, this multiplicity of solutions collapses in the spike limit $\sigma \rightarrow 0$, in which the only physical solution is \footnote{In the limit $\sigma \rightarrow 0$, equation \eqref{eq:inverted_algeraic_system} becomes $\rho'_i(c-\sum_j\rho'_j G(x_i;x_j))=0$. Since taking any $\rho'_i=0$ gives a solution of the system without the $i$th spike, these are unphysical solutions.}
\eq\bm{\rho}'=c~\bm{G}^{-1}\hat{\bm{e}}\qe
where $\bm{G}_{ij}=G(x_i;x_j)$ and $\hat{\bm{e}}$ is the column vector with all unit entries. This is precisely the same solution obtained in the perfect sink limit $\lambda \rightarrow \infty$ of the point sink system (\eqref{eq: plasmid_sol} and \eqref{eq:linear_system}). Thus, in the singular spike limit, steady-state solutions of the Turing system are equivalent to that of a system of perfect sinks, a surprising equivalence given the inverted pre-factor in \eqref{eq:outer_equation}.

We next solve the system for a single arbitrarily positioned spike. We find two solutions corresponding to different spike amplitudes
\eq \frac{\rho'_{1,\pm}}{c}=\frac{1\pm\sqrt{1-4\sigma G(x_1;x_1)}}{2G(x_1;x_1)}~.\non\qe
and corresponding masses $M_{1,\pm}=c-\rho'_\pm$. Since we only ever observe spikes within large amplitudes, i.e. patterns in which almost all the mass of the system is contained with spikes, we assume that the low amplitude solution is unphysical for finite $\sigma$ and not only in the spike limit $\sigma \rightarrow 0$ (or unstable in the context of the time-dependent system, see below). Defining the flux-differential across a spike analogously to equation \eqref{eq:flux_differential_plasmid},
\eq \Delta J_i(\bm{x})= -\frac{D_u}{2}\sum_j \rho'_j \left[ G_x(x_i^+;x_j) + G_x(x_i^-;x_j) \right],\qe
we find a linear dependence on the spike position in the regime $\kappa \ll 1$
\begin{align}
    \frac{\Delta J_1}{c\delta L} &=-\frac{1}{2}\frac{\rho'_+}{c}\frac{\sinh(2\kappa\frac{x_1}{L})}{\sinh(\kappa)}\label{eq: flux_differential_spike}\\ &\approx-\frac{\left(1+\sqrt{1-4\sigma}\right)}{2}\frac{x_1}{L}+ \mathcal{O}(\kappa^2)~,\non
\end{align}
consistent with our numerical observations (Fig. 4A,B) and just as we found for the non-inverted sinks in the previous section (Fig. 3C). Furthermore in the spike limit, $\sigma \rightarrow 0$, $\Delta J_1=c\delta x_1$ is linear in $\delta$, consistent with our numerical observations (for which $\sigma=0.0463$) (Fig. 2B). We also find that $M_{1+}$, the total mass of $u$, is minimised at mid-domain (Fig. 4B). However, while the mass and flux differential displayed very similar qualitative profiles (Fig. 4B), the agreement was not quantitative. This is likely due to the nature of the approximation and/or because our solution is not sufficiently spike-like. We will see in the next section that our main result is unaffected.

The observations that peaks in a Turing pattern move with a velocity proportional to the flux-differential across them (Fig. 2A,B, Fig. 4A, Fig. S3) suggests that the spike approximation can be extended to account for spike movement by specifying the spike velocities as
\eq \dd{x_i}{t}=\nu\Delta J_i(\bm{x})~,\label{eq:velocity_spikes}\qe
where $\nu$ is some unknown parameter. By the correspondence with point sinks, this expression becomes equivalent in the spike $\sigma \rightarrow 0$ and long diffusive length-scale $\kappa \rightarrow 0$ to a description in which the mass $M$ acts as potential and the sinks as over-damped particles as in \eqref{eq:sink_velocity_mass}. Together with \eqref{eq:inverted_algeraic_system}, \eqref{eq:velocity_spikes} defines a differential-algebraic system for the dynamics of $n$ spikes. However, not all spike configurations are stable. Based on our numerical observations, stable solutions consist only of regularly positioned spikes of the same height ($\rho'_i=\rho'$), also referred to as symmetric spike solutions, just as for the Turing system itself. Consistent with this, the flux-differentials of these solutions vanish, $\Delta J_i(\bar{\bm{x}})=0$, via the properties of the Green's function, just as for the points sinks of the previous section (and Appendix A).

As an example, we consider the case of two spikes. For our parameter set the system then has up to four real solutions for each configuration $(x_1,x_2)$. In Fig. 4D, we show sample trajectories (of the real solution branch with the smallest mass $M$; the other real solutions leads to very weak sinks and almost uniform $u(x)$). The similarity to the numerical observation (Fig. 4C) and the system of point sinks in the previous section (Fig. 3D) is apparent. In both cases, the steady-state solution consists of quarter position peaks/spikes and minimises the mass of $u$.

Let us summarise our results. We have shown that the movement and positioning of peaks in a Turing pattern is akin to that of a system of moving point sinks. Firstly, the regular, periodic steady-state positions are a result of the flow (creation, diffusion, decay) of mass through the system, and not by some dominant linear mode. The steady-state configuration is the one for which all the flux-differentials balance and this is also the configuration that minimises the total mass of the fast species. We found empirically that the movement of peaks (in the slow species) is well described by the peak velocity being proportional to the flux-differential of the fast species across it. Furthermore, in the spike limit, this is equivalent to the total mass of the fast species acting as a potential through which the peaks move as over-damped particles. While this does not imply that the mass or some other function acts as a potential away from this limit (i.e. that $\Delta \bm{J}$ is a conservative vector field in general), the mass is nonetheless minimal at the steady-state configurations (Fig. 4A,C). In the next section, we will see that we can use the stead-state mass to compare the `energy' of patterns with different numbers of peaks and in this way predict the preferred {\it number} of peaks at steady state, and not just their positions.

\section*{Competition and Pattern Selection}

\begin{figure}[!htb]
\centering
  \includegraphics[width=\linewidth]{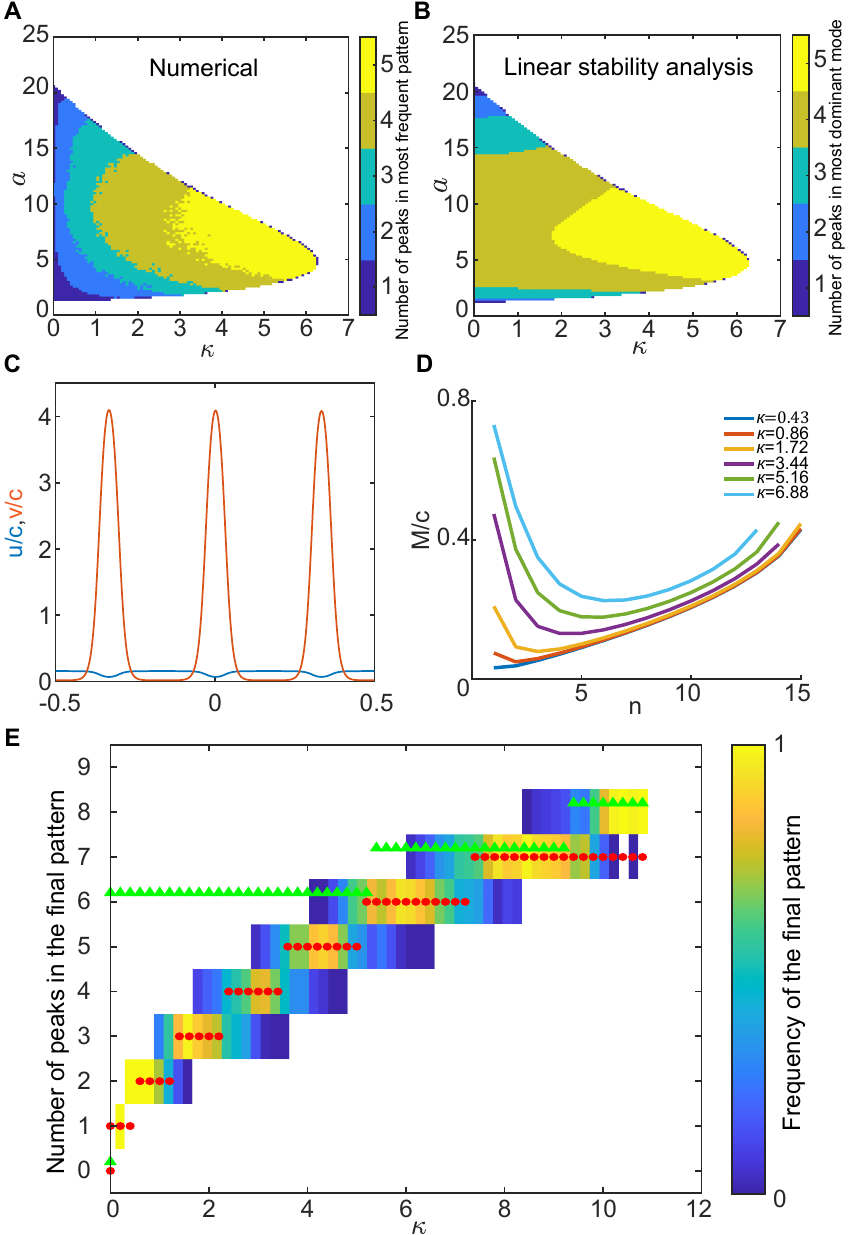}
 \caption{Mass minimisation predicts peak number in reaction-diffusion systems. (\textit{A}) The number of peaks in the most frequent steady-state pattern is plotted as a function of $a$ and $\kappa$. For each set of parameters, the most frequent pattern was obtained from $5$ simulations each initialised with a different random perturbation from the uniform state. The simulations were ran for long enough to ensure the steady-state pattern was reached. (\textit{B}) The number of peaks in the mode with the greatest growth rate as predicted by linear stability analysis is plotted as a function of $a$ and $\kappa$. Plots (A) and (B) are similar for $\kappa \gg 1$. They disagree for $\kappa \leq 1$, which indicates coarsening. (\textit{C}) Example of a steady state pattern in the spiky limit. (\textit{D})  Normalised total mass $M/c=1-n\rho'_+/c$ plotted as a function of $n$ for different values of  $\kappa$. There exists a critical $n$ for which the mass is minimal. (\textit{E}) The numerically obtained distribution of peak number at steady state for different values of $\kappa$ (colour scale) overlaid with the prediction of the dominant pattern from linear stability (green triangles) and the prediction from mass minimisation (red circles). Mass minimisation correctly predicts the number of peaks at steady state. Data from $50$ simulations for each parameter set. Parameters: Default values as in Figure 1 with $L=4$ except (\textit{C})-(\textit{E}) which use $D_v = 0.006$ (to make peaks narrower). See also Figs S4. and S5}
 \label{fig5}
\end{figure}

We have seen that in the mass-conserved limit $\delta\rightarrow0$, the model exhibits a complete coarsening effect in which the only stable patterns consist of a singe peak positioned somewhere in the interior of the domain (Fig. 2D) or, on a short domain, an interface. We have also seen that for small $\delta$ the model exhibits incomplete coarsening. With our default parameters (with $L=4$), linear stability predicts that mode $n=8$ (four peaks) will dominate (Fig. 1A). While this is true initially, the pattern subsequently coarsens so that we most frequently obtain two peaks (dominated by mode $n=4$)(Fig. 1B,C, S1D). This coarsening effect is also referred to as a competition instability and has previously been studied in the context of spike solutions \cite{Kolokolnikov2005,Kolokolnikov2006,Tzou2013}. To our knowledge, there is currently no way to determine which pattern is finally obtained. Note that this is a narrower question than asking which patterns are stable, since Turing systems are in general multi-stable.

This motivated us to explore the connection between coarsening and the flow rate $\delta$ in more detail. We measured the distribution of steady state patterns obtained for different values of $\delta$ (through the dimensionless parameter $\kappa$) and compared against the prediction of linear instability (Fig 5A,B). We used periodic boundary conditions to avoid peaks on the boundary that are not described by the spike approximation. We found that for $\kappa\gtrsim 1$ linearly stability analysis correctly predicts the dominant mode at steady state. However, for $\kappa\lesssim 1$, a coarsening process occurs and the steady-state pattern is dominated by a lower mode than that predicted. Given our previous observations on the role of the diffusive length-scale, we explain these results as follows. When the diffusive length-scale is longer than the domain size, all peaks compete for $u$ molecules created across the domain. Whereas, when the length-scale is short, peaks only absorb molecules of $u$ created within a distance given by the diffusive length-scale and therefore compete less or not at all. Competition is also exasperated by the fact that decreasing $\delta$ also decreases the total flux through the system ($c\delta L$).

We next applied the spike approximation developed in the previous section. We decreased $D_v$ from the default value so that the obtained pattern was reasonably spike-like (Fig. 5C) while at the same time not resulting in a very much enlarged Turing space (since we want to sweep over different values of $\delta$). We considered only symmetric, regularly positioned spike solutions, which are the only observed steady-state solutions. For $n$ spikes, we obtain two possible values of $\rho'$, of which we take the larger, $\rho'_+$ (the other corresponds to extremely weak spikes i.e. $\rho'_-\approx 0$). This gives a solution $u(x)=c-\rho'_+ \sum_i G(x;\bar{x}_i)$ with mass 
\eq 
M/c= 1- n\frac{1+\sqrt{1-2\kappa\sigma\coth(\frac{\kappa}{2n})}}{\kappa\coth(\frac{\kappa}{2n})}~.
\label{eq:rho_mass_min}
\qe
Note that for a real solution we must have $1>2\kappa\sigma\coth(\frac{\kappa}{2n})$. Therefore, for a given choice of parameters, there is an upper bound on the number of spikes that a solution can contain. In general, a solution exists for multiple values of $n$. However, numerically, we observe a very narrow distribution of the number of peaks (Fig. S1, Fig. 5F). We hypothesised that mass minimisation might play a role. Indeed, when we examined the mass $M$ of solutions consisting of different numbers of spikes at there respective steady-state positions, we found that the mass is minimal for a specific number of spikes (Fig. 5D). This could also be seen by plotting the mass as a function of $\kappa$ for different values of $n$ (Fig. S4). The value of $n$ at the minimum decreases with $\kappa$, with a single spike being minimal at $\kappa \rightarrow 0$. The curves invert so that as $\kappa$ is increased multiple spikes produce the lowest mass. Given that we have already shown that the mass of $u(x)$ is minimal at the steady-state, we hypothesised that it could also be used to compare solutions with different numbers of peaks and therefore identify a preferred `minimum energy' state.

We compared the number of spikes predicted by this mass minimisation principle against the distribution of patterns obtained numerically (starting from a small random perturbation around the uniform state). We found remarkable agreement (Fig. 5E, red circles). Mass minimisation correctly predicts the most frequent pattern obtained over the entire range of $\kappa$, including, most importantly, the regime in which coarsening occurs. There is significant deviation only at the transition points and close to exiting the Turing regime at high $\kappa$. In comparison, the linear prediction only agrees for the highest values of $\kappa$ i.e. close to onset (Fig. 5E, green triangles). Remarkably, the prediction was also reasonably accurate even when the solution is not very spike like, as for our default parameter set (Fig. S5) and towards the boundary of the Turing space (Fig. 2). Thus, mass minimisation not only explains {\it where} the peaks of a Turing pattern are positioned but also {\it how many} peaks there will be at steady-state, after any coarsening has taken place. Importantly, it does so far from onset and hence outside the region where weakly non-linear approaches such as the amplitude equations are valid.

\begin{figure}[htb]
\centering
  \includegraphics[width=\linewidth]{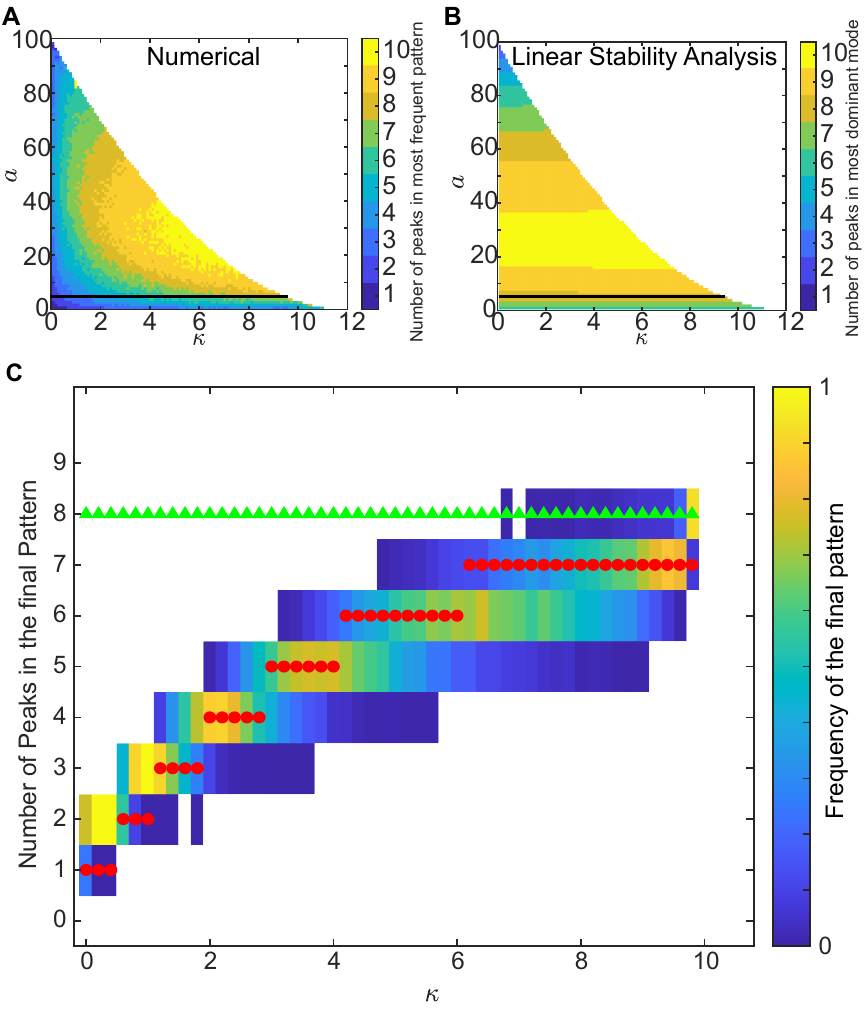}
  \caption{Mass minimisation correctly predicts the selected pattern after coarsening in the Brusselator. (\textit{A}). The number of peaks in the most frequent steady-state pattern of the Brusselator model is plotted as a function of model parameters $a$ and $\kappa$ (see Appendix C). For each set of parameters, the most frequent pattern was obtained from $10$ simulations each initialised with a different random perturbation from the uniform state. The simulations were ran for long enough to ensure the steady-state pattern was reached. (\textit{B}). The number of peaks in the mode with the greatest growth rate as predicted by linear stability analysis is plotted as a function of $a$ and $\kappa$. Coarsening observed over much of the range of $\kappa$ and the disparity with the linear stability prediction increase as $\kappa \rightarrow 0$. Parameters: $d=100$ with $D_u=1,D_v=0.01$ and $\Gamma=15000$. (\textit{C}). As Fig. 5E but for the Brusselator model. The numerically obtained distribution of the number of peaks at steady state for different values of $\kappa$ (colour scale) is overlaid with the prediction of the dominant pattern from linear stability (green triangles) and the prediction from minimisation the mass of the fast species \eqref{eq:mass_brusselator} (red circles). Mass minimisation again correctly predicts the number of peaks at steady state. Data from $500$ simulations for each parameter set. Black line in (\textit{A}) and (\textit{B}) shows the range of $\kappa$ values used within the $(a,\kappa)$ Turing space. Parameters: $d=100, a=3.75$, $\Gamma = 15000$.}
  \label{Fig6}
  \end{figure}
  
To investigate the general applicability of these results, we performed a similar analysis of the Brusselator model (Appendix C). This model also exhibits coarsening when the flow of mass through the system is slow (Fig. 6A,B). Using a similar analytic approximation as before, we obtained an expression for symmetric n-spike solutions (Appendix C). We found that the total mass of the fast species
\eq M=\frac{6n(b+1)^{3/2})}{a\sqrt{\Gamma}}+\frac{\Gamma b}{12 n^2 b(n+1)},\label{eq:mass_brusselator}\qe
is again minimised for a particular number $n=n_c$ of spikes and we found this minimum to be an excellent predictor of the final pattern obtained after coarsening (Fig. 6C). Thus mass minimisation may serve as a general principle for coarsening and pattern selection in Turing systems.

\section*{Discussion}
One of the main challenges for the physics of pattern formation is the prediction of which pattern will be obtained, not only at onset, i.e. at entry into the parameter space giving patterns, but generically for any parameter values. While linear stability analysis can give a prediction for the dominant mode, non-linear effects mean that it can be inaccurate. Furthermore, as we have argued, linear analysis cannot explain the periodic nature of final patterns nor (with reflexive boundary conditions) the positioning of peaks within the domain, which occurs dynamically in several settings that are outside of the linear regime (e.g. domain growth, coarsening, initialised peaks). Weakly non-linear approaches, such as the use of the method of amplitude equations, do exist but they are much less useful far from onset and while a recent phase-space approach has been introduced to study steady-state patterns in two-component mass-conserved systems \cite{Halatek2018b,Brauns2018}, we still lack a general theory of pattern dynamics and selection far from onset.

Here, we have presented a generalisable physical principle that can explain the positioning and preferred steady-state number of peaks in a Turing system \eqref{eq:model_equations}. To our knowledge, mass minimization is the first framework that can predict the final steady-state pattern in the presence of coarsening i.e. which of the generally several stable patterns is preferentially selected. Our expectation is that this will guide the development of new non-linear approaches for the study of pattern selection, far from onset. The insight came from analysing the behaviour of a diffusive system consisting of point sinks that move with a velocity proportional to the gradient. We showed that the flow of mass through such a system leads to sinks being positioned symmetrically and evenly spaced across the domain (regularly positioned), as this is the unique configuration for which the gradient across each sink vanishes. Importantly, we showed that this configuration also uniquely minimises the total mass in the system. Consistent with this we showed that in the long diffusive length-scale and perfect sink limit, the total mass acts as a potential and the sinks as over-damped particles.

We found that in the Turing system, peaks (of the slowly diffusing species $v$) also move toward the regularly positioned configuration, with the same dynamics as point sinks and, in doing so, minimise the total mass of the fast species $u$. In the singular limit $D_v \ll D_u$, in which the peaks of the Turing pattern become narrow point-like spikes, an analytical approximation showed that $u$ is indeed described by diffusion in the presence of point sinks but where the Dirac delta function terms have a $1/u$ pre-factor. This `inverted' sink term leads to the total mass of $u$ having a non-trivial dependence on the number of spikes and the rate of mass flow through the system. As a result, there is a critical number of spikes (and hence wavelength) that minimises the mass of $u$. We therefore hypothesised that the mass could act as a multi-well potential, i.e. that it could be used to assign, at least at the steady-state, an energy to patterns with different numbers of peaks and thereby identify the preferred steady-state. In particular, we asked whether it could predict the steady-state pattern obtained after the coarsening that occurs in our model. We found that this `mass minimisation' principle could indeed predict, almost perfectly, agreement the obtained patterns (Fig. 5E) and we confirmed this for another Turing system that exhibits coarsening (i.e. disagreement with the linear prediction), the Brusselator (Fig. 6).  In conclusion, the peaks of a Turing pattern move according to the flux-differential of the substrate species across them and both the number and positions of peaks at steady state are such that the mass of the substrate species is minimised.

The predicted number of peaks decreases with the flow of mass through the system, with the minimal mass occurring for a single peak in the mass-conserved limit. Therefore mass minimisation also provides a physical explanation for the complete coarsening down to a single peak observed in two-component mass-conserved systems \cite{Otsuji2007,Ishihara2007}. Our previous three-component mass-conserved model \cite{Murray2017}, consisting of a two-component Turing system coupled linearly to a third species, did not display such complete coarsening. The role of mass flow provides the explanation. In the three-component mass-conserved system, there is still mass flow through the Turing subsystem and the rate of flow controls the strength of the competition, similar to the current model. Thus mass-minimisation appears to be a fundamental principle behind the behaviour of Turing systems and likely pattern forming systems in general.

Are our results applicable to other systems? If mass minimisation of the fast (substrate) species is central to Turing patterning, then the mass should be minimisable. In our model, the total mass of the system $\int u+v ~dx $ at steady state is fixed. In the Gray-Scott model \cite{Pearson1993} a different positive linear combination of the two species is fixed, while in the Schnackenberg and Brusselator models it is the total mass of the slow species.  What happens then in models that have the mass of the fast species fixed? We consider the following class of systems:
\begin{align}
\partial_t u &= D_u \partial^2_x u -f(u,v)+a-u\non\\
\partial_t v &= D_v \partial^2_x v +f(u,v)+b\non
\end{align}
with $D_v<D_u$. Note that at steady state the mass of $u$ is fixed. It is straightforward to show (see Supporting Information) that a system of this form cannot admit a Turing instability for any $f$, consistent with a general principle of mass minimisation. This result holds even if we replace $u$ in the last term by any function $g(u)$ with $g'(u)>0$ at the homogeneous fixed point, and a correspondingly different measure for the mass of $u$. Thus, at least for the class of models above, the total mass of the fast species at steady state must not not generically be a fixed constant. Equivalently stated, systems in which mass leaves the system through only the fast species cannot exhibit a Turing instability.

The outer equation for $u$ obtained in the singular limit $D_v \ll D_u$ (\eqref{eq:outer_equation}), has the same form in other systems of the substrate-depletion type such as the Schnakenberg \cite{Iron2004} and Brusselator \cite{Tzou2013} and indeed both of these model exhibit the same peak movement towards regular positioning (see Supporting Information). Substrate-inhibition models that have peaks of the two species overlapping, such as that of Gierer and Meinhardt \cite{Gierer1972}, also exhibit peak movement towards regular positions. However, the outer equation of these models have a point source term rather than a point sink \cite{Iron2001}. The effect of this on mass minimisation remains to be tested.

Our results indicate that both the final position and number of peaks of a Turing pattern are such that the mass of the fast species is minimal. There is therefore minimisation with respect to $n$ continuous variables (the peak positions) and with respect to the discrete variable $n$ itself. Might the mass or some other function act as potential for the dynamics of peaks? Could it determine which peaks coarsen and when? This remains to be seen but the answer may be connected to the changes in the existence and/or stability of the different solution branches of the differential-algebraic system \eqref{eq:velocity_spikes}. Finally, unlike flux-balance, mass minimisation extends naturally beyond point sinks and to higher dimensions. It may therefore be useful in the study of more complicated structures such as the stripes, hexagons and spots that appear in two dimensions, especially far from onset.

\subsection*{Numerical Methods}
The simulations were performed in a spatial lattice $x \in [-\frac{L}{2}, \frac{L}{2}]$ and time domain $t \in [0, T]$, where $L$ is the length of the spatial domain and $T$, the total time. The MATLAB solver $pdepe$ was used to solve the time-dependent equation \eqref{eq:model_equations}. The simulations were performed with the following \textbf{default} parameters (unless explicitly stated otherwise):
\begin{align}
    \begin{split}
 D_u = 0.3, D_v = 0.012, L=2, c=300, \\\beta = 1.5 \times 10^-4, \gamma = 3.6, \delta = 0.014.
 \end{split}
\end{align}
the equivalent dimensionless parameters are
\begin{align}
  d=25, a=3.75, b= 0.0039, \Gamma = 1200.
  \label{eq: default_parameters}
\end{align} 
The simulations were ran long enough so as to obtain the true steady state. The relative and absolute tolerances in the difference between two values of iteration were set to $10^{-6}$ and $10^{-12}$ respectively. We used reflective boundary conditions
\begin{align}
    \frac{\partial u}{\partial x}\Big|_{x=-L/2,L/2} = \frac{\partial v}{\partial x}\Big|_{x=-L/2,L/2} = 0.
\end{align}
except for Figure 5 and 6 Figures S5 and S7, where we use periodic boundary conditions,
\begin{align}
\begin{split}
    u|_{x=-L/2} = u|_{x=L/2},\quad u'|_{x=-L/2} = u'|_{x=L/2}, \\
    v|_{x=-L/2} = v|_{x=L/2}, \quad v'|_{x=-L/2} = v'|_{x=L/2}.
\end{split}
\non
\end{align}
The initial conditions were taken to be a random perturbation around the homogeneous steady state (drawn from a normal distribution with
standard deviation of 1\%).

The differential algebraic systems \eqref{eq:sink_velocity}, \eqref{eq:sink_velocity_mass} and \eqref{eq:velocity_spikes} were solved using the $ode15s$ solver.

\subsection*{Numerical comparisons for Figure 4}
To compare the movement of peaks in the simulations with analytical calculations (spike approximation), we initialise the peaks at different positions (by translation of the steady state pattern) and monitor the evolution of the system. In Fig. 4A (single peak), we calculate the mass of the fast species, $M(t)=\int_{-L/2}^{L/2} u(x,t) dx$, and the flux on the peak $\Delta J_s$ (\eqref{eq: flux_differential_peak}), at each time step as the peak approaches mid-domain. In Fig. 4C (two peaks), we generated the contours of $M$ by simulating the Turing system with around 200 different initial peak positions $(x_1,x_2)$ (some of which are overlayed in white) and interpolating over the trajectories. Note that it is not clear how to define the flux differential across each peaks in the case of more than one peak. Similarly, in Fig. 4D, we solved the nonlinear algebraic system in \eqref{eq:inverted_algeraic_system} for different sink positions $(x_1,x_2)$. We obtain at most $4$ different solutions of which we take the one with the lowest total mass $M$. Several analytical trajectories were obtained by solving the differential-algebraic system in Matlab consisting of \eqref{eq:inverted_algeraic_system} and \eqref{eq:velocity_spikes}, initialised with the solution having the lowest mass.

\appendix
\section{Point sinks, flux balance and mass minimisation}

In this appendix, we prove an important result described in the main text. The solution to \eqref{eq: plasmid_eq} is given by
\begin{align}
    A(x) &= c-\sum_i \mu'_i G(x;x_i)\label{eq:A}
\end{align}
where the Green's function is defined in \eqref{eq:greens_function_equation}
and the $\mu'_i=\mu'_i(\bm{x})$ are determined by the algebraic equations
\eq \mu'_i=\lambda(c- \sum_j \mu'_j G(x_i;x_j))~.\label{eq:muprime_def}\qe

\subsection*{Properties of the Green's function}
The explicit form of the Green's function is
\begin{align}
    G(x;x_i) =\frac{\kappa}{2} \frac{\cosh{(\kappa\frac{x+x_i}{L})}+\cosh{(\kappa\frac{\mid x-x_i \mid - L}{L})}}{\sinh{(\kappa)}}~.\label{eq:greens_function}
\end{align}
The derivative of $G(x;x_i)$ with respect to $x$ is discontinuous at $x=x_i$
\begin{align}
    G_x(x;x_i) &= \begin{cases}
     \frac{\kappa^2}{2L} \frac{\sinh(\kappa\frac{x+x_i}{L})-\sinh(\kappa \frac{x_i-x-L}{L})}{\sinh(\kappa)} & -\frac{L}{2} \leq x < x_i\\
    \frac{\kappa^2}{2L} \frac{\sinh(\kappa \frac{x+x_i}{L})+\sinh(\kappa \frac{x-x_i -L}{L})}{\sinh(\kappa)} &   x_i < x \leq \frac{L}{2}~.
    \end{cases}
\non
\end{align}
Note the following property
\begin{align}
    G_x(x_i^+;x_j) - G_x(x_i^-;x_j) = -\frac{\kappa^2}{L}\delta_{ij}~.
    \label{eq: Gxdiff}
\end{align}
Using this, the flux-differential $\Delta J_i$ defined in the main text can be written as
\begin{align}
\Delta J_i &=-\frac{D}{2}\sum_j\mu'_j\left[G_x(x_i^+;x_j)+G_x(x_i^-;x_j)\right]\non\\
    &=-D\sum_j\mu'_j\left[ G_x(x_i^+;x_j) + \frac{\kappa^2}{2L}\delta_{ij}\right]~. \label{eq: flux_on_plasmid}
\end{align}

We next give some properties of the Green's function when the sinks are called regularly positioned, i.e. when they are evenly spaced across the domain with positions
\begin{align}
    \bar{x}_i = \frac{L}{n}i - \frac{L}{2}(\frac{1}{n}+1), 
    \label{eq: regular_positions}
\end{align}
where $\bar{x}_i$ is the position of $i$ th sink and $n$ is the total number of sinks.
\subsubsection*{Property I}
Evaluating the Green's function at the sink positions for regularly positioned sinks defines a symmetric matrix $G_{ij}\coloneqq G(\bar{x}_i;\bar{x}_j)$. Consider the sum of the $j$ column, 
\begin{widetext}
\begin{align}
        \sum_j G_{ij}& = \frac{\kappa}{2\sinh(\kappa)}\left[\sum_{j=1}^{n}\cosh(\kappa\frac{\bar{x}_i+\bar{x}_j}{L}) + \sum_{j =1}^{i} \cosh(\kappa\frac{\bar{x}_i-\bar{x}_j-L}{L}) + \sum_{j = i+1}^{n} \cosh(\kappa\frac{\bar{x}_j-\bar{x}_i-L}{L}) \right] \non\\
        &= \frac{\kappa}{2\sinh(\kappa)}\left[\sum_{j=1}^{n}\cosh(a(i+j-1-n))+\sum_{j =1}^{i} \cosh(a(i-j-n)) + \sum_{j = i+1}^{n} \cosh(a(j-i-n))\right]\non\\
        &= \frac{\kappa}{2}\coth(\frac{\kappa}{2n})\quad\forall ~i, 
        \label{eq:green_func_property}
\end{align}
\end{widetext}
where $a=\frac{\kappa}{n}$ and the last step follows from the identity,
\begin{align}
    \sum_{j=1}^{n} \cosh(a(j+m)) = \csch(\frac{a}{2})\sinh(\frac{an}{2})\cosh(\frac{a}{2}(2m+n+1)). \non
\end{align}
We can similarly define a matrix $G^+_x$ by evaluating the derivative of the Green's function at regular positioning $(G_x^+)_{ij}=G_x(\bar{x}^+_i;\bar{x}_j)$.
Summing over the $j$ th column we find
\begin{align}
        \sum_j (G^+_x)_{ij}
    &=-\frac{\kappa^2 }{2L}
\label{eq:sumGx}
\end{align}
by using the identity, 
\begin{align}
    \sum_{j=1}^{n} \sinh(a(j+m)) = \csch(\frac{a}{2})\sinh(\frac{an}{2})\sinh(\frac{a}{2}(2m+n+1)) .\non
\end{align}

\subsubsection*{Property II}
Since the summation of $G$ over any of its rows or columns is the same, the vector of $1$s,  $\boldsymbol{\hat{e}}$, is an eigenvector of $G$.
Evaluating the defining equations for the $\bm{\mu'}$, equation \eqref{eq:muprime_def}, at regular positioning $\bm{x}=\bar{\bm{x}}$, we obtain the matrix equation
\begin{equation}
    (\lambda G + 1) \boldsymbol{\mu'}(\bm{\bar{x}}) = \lambda c \boldsymbol{\hat{e}}~.
    \label{eq: b_matrix_eq}
\end{equation}
Since $\boldsymbol{\hat{e}}$ is an eigenvector of $\lambda G+1$, we must have that
\begin{align}
    \boldsymbol{\mu'}(\bm{\bar{x}}) = C_1\boldsymbol{\hat{e}}~, \non
\end{align}
i.e. all the $\mu'_i$ are identical at regular positioning or in other words, the profile of $A$ is symmetric. We can sum over any row and use \eqref{eq:green_func_property} to find
\eq \mu'_i(\bm{\bar{x}})=\frac{\lambda c }{1+\lambda\frac{\kappa}{2}\coth{(\frac{\kappa}{2n})}} ~. \label{eq: bprime_regular}\qe

\subsection*{Regular positioning and flux balance}
The flux differential across each sink is given by,
\begin{align}
\Delta J_i &=-D\sum_j\mu'_j\left[ G_x(x_i^+;x_j) + \frac{\kappa^2}{2L}\delta_{ij}\right]~.
\label{eq: flux_on_sink}
\end{align}
We evaluate this expression for regularly positioned sinks $\bm{x}=\bar{\bm{x}}$. First we know from equation \eqref{eq: bprime_regular} that all $\mu_j'$ are identical for regularly positioned sinks. Then from equation \eqref{eq:sumGx}, it follows that immediately that the flux-differentials vanish at regular positioning
\begin{align}
    \Delta J_i(\bar{\bm{x}}) = 0\quad\forall ~i ~.
 \label{eq: flux_regular_position}
\end{align}

To show that the regularly positioned configuration is the unique configuration for which the flux-differentials vanish, we perform a power series expansion of $\Delta J_i$ in $\kappa$. It then suffices to show uniqueness for the $\kappa^0$ term. We first expand $\mu'_i$ and $G'(x_i;x_j)$
\begin{align}
    \mu'_i &= \mu_{0i} + \mu_{2i} \kappa^2 + ... \non\\
    G(x_i;x_j) &= G_0(x_i;x_j) + G_2(x_i;x_j)\kappa^2 + ...\non
\end{align}
For the lowest order terms, we find first that $G_0(x_i;x_j)=1$. Inserting this into the defining equation for the $\mu'_i$ (\eqref{eq:muprime_def}), we have
\eq \mu'_{0i} = (c-\sum_j \mu '_{0j})\lambda\quad\forall ~i \non\qe
which has solution, \eq\mu'_{0i}=\mu'_0=\frac{\lambda c}{1+n\lambda}.\label{eq: mu0}\qe
We then have
\begin{widetext}
\begin{align}
\frac{\Delta J_i}{\delta L}  &= -\frac{D}{\delta L}\sum_{j}\mu'_j \left[ G_x(x_i^+;x_j) + \frac{\kappa^2}{2L}\delta_{ij} \right]\non \\
    &=-\frac{1}{2} \left[\sum_{j=1}^{n}\mu'_j \frac{\sinh(\kappa\frac{x_i+x_j}{L})}{\sinh(\kappa)} + \sum_{j =1}^{i} \mu'_j \frac{\sinh(\kappa \frac{x_i-x_j-L}{L})}{\sinh(\kappa)}- \sum_{j=i+1}^{n} \mu'_j\frac{\sinh(\kappa \frac{x_j-x_i-L}{L})}{\sinh(\kappa)}+\mu'_i \right] \non\\
   &=-\frac{\mu_0'}{2L} \left[\sum_{j=1}^{n} (x_i+x_j) + \sum_{j=1}^{i-1} (x_i-x_j - L) - \sum_{j=i+1}^{n} (x_j-x_i-L) \right]  +O(\kappa^2)\non\\
    &= -\frac{n\mu_0'}{L}\left[x_i -\frac{L}{n}i + \frac{L}{2}(\frac{1}{n}+1)\right]+O(\kappa^2)~.
\label{eq: flux}
\end{align}
\end{widetext}

Hence, all the flux-differentials $\Delta J_i$ vanish uniquely for regularly positioned sinks $x_i=\bar{x}_i=\frac{L}{n}i-\frac{L}{2}(\frac{1}{n}+1)$. If we add time-dependence to the system by specifying the sink velocities as being proportional to the their flux-differential
\eq \dd{x_i}{t}=\nu \Delta J_i(\bm{x})~,\label{eq:sink_movement}\qe
then regularly positioned sinks is the unique fixed point of the resultant dynamical system (specified by the differential-algebraic system defined by equations \eqref{eq:muprime_def}, \eqref{eq: flux_on_plasmid} and \eqref{eq:sink_movement}. Given that the domain is bounded, the fixed point is stable.

\subsection*{Regular positioning and mass minimisation}
The total mass (or rather concentration) of $A$ is readily given by integrating equation \eqref{eq:A}
\eq M(\bm{x})=\frac{1}{L}\int_{-\frac{L}{2}}^{\frac{L}{2}} A(x) dx=c-\sum_{i=1}^n\mu'_i \label{eq: total_mass}~.\qe
We would like to show that the regularly positioned configuration is the unique stationary point of $M$. First, we will show that
\begin{align}
\left.\pdd{}{x_m}M\right|_{\bm{x}=\bm{\bar{x}}}=  -\left.\frac{\partial}{\partial x_m}\sum_i \mu'_i\right|_{\bm{x}=\bm{\bar{x}}} =0~.
    \non
\end{align}
Using \eqref{eq:muprime_def} ($\mu'_i=\lambda(c- \sum_j \mu'_j G(x_i;x_j))$), we can evaluate the derivative of $\sum_i\mu_i'$ with respect to an arbitrary sink position $x_m$,
\begin{align}
    \frac{\partial }{\partial x_m}\sum_i \mu'_i  &= -\lambda\sum_{i,j} \mu'_j G_{x_m}(x_i;x_j) -\lambda \sum_{i,j} G(x_i;x_j)\frac{\partial}{\partial x_m}\mu'_j,
\non
\end{align}
Evaluating this expression at regular positioning, and defining $C :=\sum_j G(\bar{x}_i;\bar{x}_j) = \frac{\kappa}{2}\coth (\frac{\kappa}{2n})$ from equation \eqref{eq:green_func_property}, we obtain
\begin{align}
\left.(\frac{1}{\lambda}+C)\frac{\partial }{\partial x_m}\sum_i \mu'_i \right|_{\bm{x}=\bm{\bar{x}}} &= -\left.\sum_{i,j} \mu'_j G_{x_m}(x_i;x_j)\right|_{\bm{x}=\bm{\bar{x}}} .\non
\end{align}
We have already seen in equation \eqref{eq: bprime_regular} that all the $\mu'_j$ are identical at regular positioning. Hence, we need only evaluate $\left.\sum_{i,j}G_{x_m}(x_i;x_j)\right|_{\bm{x}=\bm{\bar{x}}}$.
Inserting the definition of $G(x_i;x_j)$ from equation \eqref{eq:greens_function} we have
\begin{widetext}
\begin{align}
   \left.\sum_{i,j}G_{x_m}(x_i;x_j)\right|_{\bm{x}=\bm{\bar{x}}}=& \left.\frac{\kappa}{2\sinh(\kappa)}\sum_{i,j}\pdd{}{x_m}\left[\cosh(\kappa \frac{x_i+x_j}{L})+\cosh(\kappa\frac{|x_i-x_j|-L}{L})\right]\right|_{\bm{x}=\bm{\bar{x}}}\non\\
    =&\frac{\kappa^2}{2L\sinh(\kappa)}\sum_{i,j}\left[\delta_{mi}\left(\sinh(\kappa\frac{x_i+x_j}{L})+sgn(x_i-x_j)\sinh(\kappa\frac{|x_i-x_j|-L}{L})\right)\right.\non\\ &\qquad\qquad\qquad\left.+\left.\delta_{mj}\left(\sinh(\kappa\frac{x_i+x_j}{L})-sgn(x_i-x_j)\sinh(\kappa\frac{|x_i-x_j|-L}{L})\right)\right]\right|_{\bm{x}=\bm{\bar{x}}}\non\\
    =&\left.\frac{\kappa^2}{L\sinh(\kappa)}\sum_{i}\left[ \sinh(\kappa\frac{x_m+x_i}{L})+sgn(x_m-x_i)\sinh(\kappa\frac{|x_m-x_i|-L}{L}) \right]\right|_{\bm{x}=\bm{\bar{x}}}\non \\
    =& \frac{\kappa^2}{L\sinh(\kappa)}\left[ \sum_{i=1}^n\sinh(a(m+i-n-1))+\sum_{i=1}^{m-1}\sinh(a(m-i-n))-\sum_{i=m+1}^{n}\sinh(a(i-m-n)) \right]\non\\
    =&\frac{\kappa^2}{L\sinh(\kappa)}\left[-\sinh(\kappa)-\sinh(-\kappa)\right]=0\non
\end{align}
\end{widetext}
where the last line follows from noting that the summations are the same is in equation \eqref{eq:sumGx} but without the $i=m$ term. We have therefore shown that regular positioning is a stationary configuration of the total mass
\eq \left.\pdd{}{x_m}M\right|_{\bm{x}=\bm{\bar{x}}} =0 ~.\non \qe
To show that regular positioning is the unique stationary point, we proceed as in the previous section and perform a power series expansion of $M$,
\eq M=M_0+M_2 \kappa^2 +\cdots \qe
It then suffices to show uniqueness for the first non-trivial order in the expansion.
For the Green's function we have
\begin{align}
    G_0(x_i;x_j) &= 1, \quad G_2(x_i;x_j) = \frac{x_i^2 + x_j^2 -  L|x_i-x_j | }{2L^2} + \frac{1}{12}.\non
\end{align}
We already saw that $\mu'_{0i}= \mu'_0=\frac{\lambda c}{1+n\lambda}$ and hence $M_0$ is a constant. Inserting these into the equation for $\mu'_{2i}$ using \eqref{eq:muprime_def},
\begin{align}
    \mu'_{2i} &= -\lambda\sum_j (\mu'_{2j}  + \mu'_{0j} G_2(x_i;x_j)),
 \label{eq: mu_second_order}
\end{align}
we obtain
\begin{align}
M_2 &=-\sum_i \mu'_{2i} \non\\&= \frac{\lambda}{1+n\lambda}\mu'_0\sum_i\sum_j  \left(\frac{x_i^2 + x_j^2 - L|x_i-x_j |}{2L^2} + \frac{1}{12}\right)~.\label{eq: M2}
\end{align}
The derivative of $M_2$ is then proportional to
\begin{widetext}
\begin{align}
    \frac{\partial}{\partial x_m}& \sum_i \sum_j \left[x_i^2+ x_j^2 - L|x_i - x_j| \right]\non \\
    &= \frac{\partial}{\partial x_m}\left[ 2n\sum_i x_i^2 - \sum_i\sum_{j\leq i} (x_i - x_j)  L - \sum_i\sum_{j>i} (x_j - x_i)  L\right]\non\\
    &= \frac{\partial}{\partial x_m}\left[ 2n\sum_i x_i^2 - L(\sum_i i x_i  - \sum_i (n -i)x_i - \sum_i \sum_{j\leq i}x_j  + \sum_i \sum_{j>i}x_j)\right] \non\\
    &= \left[4nx_m-L(m-(n-m)-(n-m+1)+m-1)\right]\non\\
    &= 4n\left[x_m - \frac{L}{n}m+\frac{L}{2}+\frac{L}{2n})\right]
\label{eq: mass_derivative}
\end{align}
\end{widetext}
which vanishes only for the regularly positioned configuration
\begin{align}
    x_m = \frac{L}{n}m-\frac{L}{2}(\frac{1}{n}+1).\non
\end{align}
Hence, we have shown that regular positioned sinks is the unique configuration for which the total mass, $M$, is stationary. Based on our numerical results, we assume that this configuration is generically a minimum.

\section{The mass acts as a potential in the limit $k\rightarrow0,~\lambda \rightarrow\infty$}
\subsubsection*{Dynamics of a single point sink}
We consider the regime in which the timescale of gradient formation is much faster than that of sink movement. The profile $A(x)$ can therefore be approximated by \eqref{eq:A}. The flux differential on a single sink is given by \eqref{eq: flux_on_sink},
\begin{align}
    \Delta J_1 &= -D \mu'_1 \left[ G_x (x_1^+;x_1) + \frac{\kappa^2}{2L}\right], \non\\
    &= -D c\frac{\lambda}{\lambda G(x_1;x_1) + 1}\left[ \frac{\kappa^2 \sinh (2\kappa x_1/L)}{2L \sinh(\kappa)}\right], 
\end{align}
where
\eq \mu_1' = \frac{\lambda c}{\lambda G(x_1;x_1)+1},\non \qe \eq\qquad G(x_1;x_1) = \frac{\kappa}{2}\left[\frac{\cosh (\frac{2\kappa x_1}{L})}{\sinh (\kappa)} + \coth (\kappa)\right]. \non\qe
Now consider the derivative of the total mass \eqref{eq: total_mass},
\begin{align}
        \pdd{M}{x_1} &= \frac{d}{dx_1}(c-\mu'_1)  = -\frac{d}{dx_1}\mu'_1 \non\\
        &= -c\left(\frac{\lambda}{\lambda G(x_1;x_1)+1}\right)^2 \frac{\kappa^2\sinh{(2\kappa x_1/L)}}{L\sinh(\kappa)}~.
\end{align}
In the limit $\kappa \rightarrow 0$ we can see that easily that $G(x_1;x_1)=1$. If sinks are also much stronger than the background decay rate i.e. $\lambda \gg 1$, then we obtain the following relation,
\eq \Delta J = -\frac{D}{2}\pdd{M}{x_1}~.\non \qe
Hence, in that limit the velocity of a single point sink can be written equivalently in terms of the derivative of total mass $M$ as,
\begin{align}
    \dd{x_1}{t} &= \nu\Delta J_1\non\\
    &=-\nu\frac{D}{2} \pdd{M}{x_1}~.
\end{align}
In the next subsection, we generalise this relation to $n$ arbitrary sinks.
\subsubsection*{Dynamics of n sinks}
We consider again the limit $\kappa \ll 1$. Then, the flux differential up to lowest order in $\kappa$ is given by \eqref{eq: flux},
\begin{align}
\begin{split}
    \frac{\Delta J_i}{c \delta L} &= -\frac{n}{L}\frac{\lambda }{1+n\lambda}\left[x_i - \frac{L}{n}i + \frac{L}{2}(\frac{1}{n}+1)\right] + O(k^2),
    \label{eq: flux_lowest_order}
\end{split}
\end{align}
where we have used the expression for the lowest order term of $\mu'$ from \eqref{eq: mu0}, 
\eq \mu'_0 = \frac{\lambda c}{1+n \lambda}.\non  \qe The derivative of $M$ up to second order in $\kappa$ from \eqref{eq: mass_derivative},
\begin{align}
\begin{split}
\frac{D}{c\delta L}\pdd{M}{x_i} &= \left(\frac{\lambda }{1+n\lambda}\right)^2 \pdd{}{x_i}\sum_{i,j}G_2(x_i;x_j) + O(k^4),\\
     &= \frac{2n}{L}\left(\frac{\lambda }{1+n\lambda}\right)^2\left[x_i - \frac{L}{n}i + \frac{L}{2}(\frac{1}{n}+1)\right]+O(k^4)
\end{split}
\end{align}
where we have used the second order expansion of $G(x_i;x_j)$ in $\kappa$ (\eqref{eq: mass_derivative}). Taken together the above equations in the limit $\lambda \gg 1$ imply,
\begin{align}
\begin{split}
   \Delta J_i = -\frac{1}{2}nD\pdd{M}{x_i}~.
   \label{eq: mass_versus_flux_dynamics}
\end{split}
\end{align}
Therefore in that limit the dynamics of the point sinks are equivalently specified by 
\begin{align}
    \dd{x_1}{t} = -\nu \frac{n}{2}D \pdd{M}{x_i} ~.
    \label{eq: mass_potential}
\end{align}
Hence for large diffusive length scale $\kappa \ll 1$ and strong sinks $\lambda \gg 1$, the dynamics of the system is akin to damped particles moving in a potential specified by the total mass. It should be noted that away from this limit, while the dynamics may be different, the two prescriptions share the same stationary points. However, for our parameter choices we found very good agreement between the two systems (Fig. 3D).

\section{Peak Movement and Pattern selection in the Brusselator}

We now present an analysis of Brusselator and show numerically and analytically that most of the features of the exploratory model in  \eqref{eq:model_equations} still holds.

The general spatial version of Brusselator \cite{Prigogine1968} is described by the following equations,
\begin{align}
\begin{split}
      \frac{\partial u}{\partial t} &= D_u \frac{\partial^2 u}{\partial x^2} - \beta uv^2 + \gamma v  \\
    \frac{\partial v}{\partial t} &= D_v \frac{\partial^2 v}{\partial x^2} + \beta uv^2 - \gamma v  + \delta c -\delta v~.
\end{split}
\end{align}
In the absence of diffusion it has a single fixed point. The model also  has the form of a mass-conserving Turing system with additional linear terms. However, note that rather than total mass being fixed at steady state, it is the mass of $v$ that is fixed
\eq \frac{1}{L}\int \bar{v} ~dx = c ~.\qe 
As before, by writing the source term as $\delta c$ we can change the turnover $\delta$ without affecting the steady state concentration of $v$. 
We non-dimensionalise the system by
\begin{align*}
   x\to \frac{x}{L}, t \to \frac{D_v t}{L^2},u \to \frac{u}{c}, v \to \frac{v}{c}
\end{align*}
to obtain
\begin{subequations}
\begin{align}
    \frac{\partial u}{\partial t} &= d \frac{\partial^2 u}{\partial x^2} + \Gamma \Big( -  auv^2 + v \Big)\label{eq: brusselator_a} \\
    \frac{\partial v}{\partial t} &=   \frac{\partial^2 v}{\partial x^2} + \Gamma \Big(    auv^2 - v + b(1-v) \Big)
    \label{eq: brusselator_b}
\end{align}
   \label{eq: brusselator}
\end{subequations}
where
\begin{align*}
    \Gamma = \frac{\gamma L^2}{D_v}, d = \frac{D_u}{D_v}, a = \frac{\beta c^2}{\gamma}, b=\frac{\delta}{\gamma}~.
\end{align*}
The fixed point is,
\begin{align}
    u_0 = \frac{1}{a}, v_0 = 1 ~.\non
\end{align}
The Jacobian is given by,
\begin{align}
J_{(u_0,v_0)} = \Gamma\begin{bmatrix}
-av_0^2 & -2au_0v_0 +1 \\
av_0^2 & 2au_0v_0-1-b
\end{bmatrix} &= \Gamma\begin{bmatrix}
-a & -1 \\
a & 1-b
\end{bmatrix}~. \non
\end{align}
The Jacobian (and hence the dispersion relation) becomes independent of $b$ for $b<<1$. Hence, we can change $b$ without significantly affecting the linear behaviour of the model.
The trace and determinant of the Jacobian are easily found to be,
\begin{align*}
    \begin{split}
        TrJ &= \Gamma(1-a-b)\\
        DetJ &= \Gamma^2ab~.
        \end{split}
\end{align*}
For the homogeneous fixed point to be stable in the absence of diffusion we also need $TrJ<0$ and $DetJ>0$. Hence, we require $a+b>1$. The Turing condition for the Brusselator is given by,
\begin{align}
  d (1-b) -a -2\sqrt{dab} >0. \non
\end{align}
We numerically solve this system, using reflexive boundary condition, by perturbing the homogeneous state as described in numerical method. Like in our model, and every Turing model we are aware of, the interior peaks of a pattern are periodic and regularly positioned. Furthermore, consistent with our results, a single peak moves exponentially to mid-domain (Figure S6A). The rate of movement was found to be proportional to $b$, or equivalently, $\delta$, the turnover rate and for $b=0$ no peak movement is observed.

In Figure \ref{Fig6}A we switch to periodic boundary conditions as in Figure 5 and compare the number of peaks in the dominant mode as predicted by the linear dispersion relation and the number of peaks in the most frequent pattern as obtained from the numerical simulations. As in the main text, we replace $b$ by $\kappa=\sqrt{\frac{b\Gamma}{d}}$, the ratio of the length of the domain to the diffusive length-scale, which we find to be a more physical parameter (the dimensionless variables are then $\Gamma$, $d$, $a$ and $\kappa$). We observed a similar coarsening behaviour as the turnover rate is decreased (and the diffusive length-scale lengthened) as in the exploratory model (Figure 5A,B), where the number of peaks in the final pattern is fewer than what is predicted by linear stability analysis for lower values of $\kappa$.

\subsection*{Mass minimization predicts the pattern obtained after coarsening}
In this section, we derive an analytical expression for the mass of the fast species in the Brusselator model in the spike limit. Let us consider the dimensionless form in \eqref{eq: brusselator}. As in the case of our toy model we consider the limit $D_v \ll Du$, where solutions of $v$ consist of narrow large-amplitude spikes, defined to have width $\epsilon$. Away from the spikes $v$ is a spatial constant $v_{out}$. Since $\int_{-1/2}^{1/2} v dx=1$, inside the spikes we have $v \gg 1$ and outside $v=v_{out}\ll 1$. We search for steady state solutions consisting of $n$ spikes at $x_i = \{ x_1,x_2,...,x_n\}$ and assume that $u$ changes slowly within each spike and can be approximated by a constant $u_i$. We shift to an inner coordinate $y = \epsilon^{-1} (x-x_i)$, within each spike. The equation for the inner variable $v_i(y)$ becomes,
\begin{align}
    \frac{1}{\epsilon^2}\frac{d^2 v_i}{d y_i^2} + \Gamma(au_i v^2-(b+1)v) = 0 \non, \\
    v_i \to 0 \text{ as } y_i \to \pm \infty \non,
\end{align}
where we have neglected the constant term since $v_i\gg 1$. The solution to this equation is
\begin{align}
    v_i = \frac{3}{2}\frac{b+1}{au_i}\text{sech}^2(\frac{\sqrt{b+1}}{2}\frac{\epsilon y_i}{2}) .\non
\end{align}
In the outer region the equation each spike is approximated by a Dirac delta function 
\eq v=v_{out}+\sum_i w_{i_1} \delta(x-x_i)~,\label{eq:v_brusselator} \qe
where $w_{i_1}$ is the weight
\eq w_{i_1} = \epsilon\int_{-\infty}^{\infty} v_i(y_i)dy_i = \frac{6}{au_i}\sqrt{\frac{b+1}{\Gamma}}~.\non \qe
The $\epsilon$ pre-factor is the spike width. To write the outer equation for $u$, we  also need the weight of the $uv^2$ term
\eq
    w_{i_2} = \epsilon a u_i\int_{-\infty}^{\infty} v^2_i(y_i)dy_i = \frac{6}{au_i} \frac{(b+1)^{3/2}}{\sqrt{\Gamma}}~.
\qe
The spike contribution from the $-uv^2+v$ is then
\begin{align}
    -w_{i1}+w_{i2} = \frac{6b}{au_i}\sqrt{\frac{b+1}{\Gamma}}~.
\end{align}
Finally since $v_{out}\ll1$, we neglect the $uv^2_{out}$ term. The outer equation for $u$ becomes,
\begin{align}
    d \frac{d^2 u}{dx^2} - \Gamma \left(  \sum_{i=1}^{n}\frac{6b}{au_i}\sqrt{\frac{(b+1)}{\Gamma}}\delta(x-x_i) \right) + \Gamma v_{out} = 0~,\non\\ \qquad -1/2 < x < 1/2, ~~u_x(\pm 1/2) = 0~.
    \label{eq: u_outer}
\end{align}
Integrating the above equation over the whole domain we obtain,
\begin{align}
\sum_i \frac{1}{u_i} &= \frac{a v_{out}}{6b}\sqrt{\frac{\Gamma}{b+1}}~.
\label{eq: ui}
\end{align}
In the outer region, $v$ appears almost flat, hence we can approximate \eqref{eq: brusselator_b} as,
\eq auv_{out}^2 - (b+1)v_{out} +b = 0~.\non\qe This implies $v_{out} = \frac{b}{b+1}$, where we ignore the cubic term as $v_{out} \ll 1$. The expression is alternatively obtained by using \eqref{eq:v_brusselator} the integral condition $\int v dx=1$. 
The solution to the outer equation now is,
\begin{align}
    u (x) = \bar{u} + \frac{6b}{a}\sqrt{\frac{(b+1)}{\Gamma}}\sum_{j=1}^{n} \frac{1}{u_j} G(x;x_j),
    \label{eq: brusselator_solution}
\end{align}
where $\bar{u}$ is a constant and the Green's function $G(x;x_j)$ is the solution to
\begin{align}
    \frac{D}{2} G_{xx}(x;x_j) + 1 &= \delta (x-x_j),\quad -1/2<x<1/2, \non\\\quad G_x(\pm 1/2) &= 0, \quad \int_{-1/2}^{1/2}G(x;x_j)dx = 0 \non
\end{align}
 given by
\begin{align}
    G(x;x_j) = -\frac{1}{D} (x^2 + x_j^2) + \frac{1}{D}\mid x - x_j\mid - \frac{1}{6D}. \non
\end{align}
We identify the coefficient $D$ in the above equations as,  $D=\frac{2d}{\Gamma}$. The constant $\bar{u}$ in \eqref{eq: brusselator_solution} is determined by the condition for $u_i$,
\begin{align}
    u(x_i)=u_i= \bar{u} + \frac{6b}{a}\sqrt{\frac{(b+1)}{\Gamma}}\sum_{j=1}^{n} \frac{1}{u_j} G(x_i;x_j). \non
\end{align}
Now we consider the situation where the spikes are symmetric ($u_i=u_c$) and regularly positioned, 
\begin{align}
    \bar{x}_i = \frac{L}{n}i - \frac{L}{2}(\frac{1}{n}+1) \non
\end{align}
as in the analysis in the main text. For convenience, we also use reflexive rather than periodic conditions as the solutions only differ by an arbitrary phase. From \eqref{eq: ui}, we then have
\begin{align}
    u_c = \frac{6}{a}\frac{(b+1)^{3/2}}{\sqrt{\Gamma}}n. \non
\end{align}
The equation for $u(x_i)$ becomes,
\begin{align}
    u(\bar{x_i}) = u_c = \bar{u} + \frac{6b}{a}\sqrt{\frac{(b+1)}{\Gamma}}\frac{1}{u_c}\sum_{j=1}^{n} G(\bar{x_i};\bar{x_j}). \non
\end{align}
Evaluating the sum in the above equation we find an expression independent of $i$,
\begin{align}
   \sum_{j=1}^{n} G(\bar{x_i};\bar{x_j}) = -\frac{1}{6Dn}\non
\end{align}
and determine that the constant $\bar{u}$ which is in fact the total mass $M$ of the fast species
\begin{widetext}
\begin{align}
    M = \int_{-1/2}^{1/2} u(x) dx &= \int_{-1/2}^{1/2}\bar{u} ~dx =  \frac{6n(b+1)^{3/2}}{a\sqrt{\Gamma}} + \frac{1}{n^2}\frac{\Gamma b}{12d  (b+1)},
\end{align}
where we have used the Green's function property $\int_{-1/2}^{1/2}G(x;x_j) = 0$. Note $M$ is minimal for a critical number of peaks $n=n_c$.

In Figure \ref{Fig6}C, we compare the number of peaks of the final pattern obtained after coarsening (black line in Figure \ref{Fig6}A) with $n_c$ and the linear prediction. We find that, like in the model of the main text, $n_c$ is an excellent predictor of number of peaks in the steady state pattern.
\end{widetext}

\bibliography{library}

\end{document}